\documentclass[%
 reprint,
 amsmath,amssymb,
 prb,
 superscriptaddress,
floatfix,showkeys,
nobibnotes
]{revtex4-1}
\usepackage{graphicx}
\usepackage{dcolumn}
\usepackage{bm}
\usepackage{color}
\usepackage{wrapfig}
\usepackage{physics}
\usepackage{fancyhdr}
\usepackage{lastpage}
\usepackage{ragged2e}
\usepackage{siunitx}
\usepackage{xcolor}
\usepackage{soul}

\usepackage{caption}
\newcommand{\be}{\begin{equation}}
\newcommand{\ee}[1]{\label{#1} \end{equation}}
\newcommand{\TU}{\textup}

\def\_#1{\textsubscript{#1}}
\def\^#1{\textsuperscript{#1}}

\begin{document}

\title{Electrostatic Steering of Thermal Emission with Active Metasurface Control of Delocalized Modes}

\author{Joel Siegel}
\thanks{J.S. and S. K. contributed equally to this work.}%
\affiliation{Department of Physics, University of Wisconsin-Madison, Madison WI 53706 USA}%

\author{Shinho Kim}
\thanks{J.S. and S. K. contributed equally to this work.}%
\affiliation{%
School of Electrical Engineering, Korea Advanced Institute of Science and Technology, Daejeon 34141, Korea
}%

\author{Margaret Fortman}
\affiliation{Department of Physics, University of Wisconsin-Madison, Madison WI 53706 USA}

\author{Chenghao Wan}
\affiliation{Department of Electrical and Computer Engineering, University of Wisconsin-Madison, Madison WI 53706 USA}

\author{Mikhail A. Kats}
\affiliation{Department of Electrical and Computer Engineering, University of Wisconsin-Madison, Madison WI 53706 USA}

\author{Phillip W. C. Hon}
\affiliation{Northrop Grumman Corporation, Redondo Beach, California 90278, USA}%

\author{Luke Sweatlock}
\affiliation{Northrop Grumman Corporation, Redondo Beach, California 90278, USA}%

\author{Min Seok Jang}
\thanks{jang.minseok@kaist.ac.kr \\ vbrar@wisc.edu}%
\affiliation{%
School of Electrical Engineering, Korea Advanced Institute of Science and Technology, Daejeon 34141, Korea
}%

\author{Victor Watson Brar}
\thanks{jang.minseok@kaist.ac.kr \\ vbrar@wisc.edu}%
\affiliation{Department of Physics, University of Wisconsin-Madison, Madison WI 53706 USA}%

\begin{abstract}
\textbf{\abstractname}
We theoretically describe and experimentally demonstrate a graphene-integrated metasurface structure that enables electrically-tunable directional control of thermal emission.  This device consists of a dielectric spacer that acts as a Fabry-Perot resonator supporting long-range delocalized modes bounded on one side by an electrostatically tunable metal-graphene metasurface.  By varying the Fermi level of the graphene, the accumulated phase of the Fabry-Perot mode is shifted, which changes the direction of absorption and emission at a fixed frequency. We directly measure the frequency- and angle-dependent emissivity of the thermal emission from a fabricated device heated to \SI{250}{\degreeCelsius}. Our results show that electrostatic control allows the thermal emission at \SI{6.61}{\um} to be continuously steered over \ang{16}, with a peak emissivity maintained above \num{0.9}. We analyze the dynamic behavior of the thermal emission steerer theoretically using a Fano interference model, and use the model to design optimized thermal steerer structures. 

\end{abstract}


\maketitle

\section*{Introduction}

The mid infrared (MIR) is an important band for applications ranging from free-space laser communications\cite{Free_Space_Optical_Paul} to chemical sensing \cite{MIR_Chemical_Sensing_Haas, Graphene_Chemical_Sensing_Rodrigo}. An optimal MIR source for these applications would be narrowband, and also offer high speed directional control, such that the beam can be rastered over a range of angles, or have a controllable focal point. Typically, such beam-steering is achieved by reflecting a beam using mechanical devices such as gimbal-mounted mirrors\cite{Mems_Beamsteering_Byung}, optical phased arrays of antenna\cite{Phased_Array_Sun, Complete_Complex_Han}, or liquid crystal-based devices\cite{Liquid_Crystal_Steering_Jesse}. While each of these techniques have their own set of advantages and disadvantages, one limitation common to them all is that they require an external source of light, such as a quantum cascade laser. 

An alternative source of MIR light is one that can be found everywhere, thermal radiation. Any material at a non-zero temperature will emit radiation over a broad range of frequencies which, at moderate temperatures (\num{0}-\SI{700}{\degreeCelsius}), is peaked in the MIR.  Though thermal emission is typically viewed as incoherent, isotropic, and broadband, recent advances in nanoengineering have demonstrated that it is possible to engineer the emissivity of a structured material to create narrowband\cite{HighQ_Zhiyu} directional\cite{Coherent_emission_Greffet} emissions that exhibits coherence. These include metasurfaces composed of non-interacting, localized resonator elements tuned to specific wavelengths, such as metallic nanoantennas\cite{thermal_emission_Padilla} or semiconducting nanostructures that exhibit sharp quasi bound-state-int the continuum resonances\cite{doiron2019non,sun2021ultra}.  To achieve coherent directional emission, meanwhile, structures that support long-range delocalized modes can be utilized. These include surface waves that are out-coupled via gratings\cite{Coherent_emission_Greffet,tsai2006high,ghanekar2022method}, Fabry-Perot (F-P) cavities\cite{Fabry_Perot_Wang}, photonic crystals\cite{lee2005coherent,celanovic2005resonant,battula2006monochromatic}, epsilon near zero modes\cite{xu2021broadband} and delocalized modes formed by coupled resonators\cite{overvig2021thermal,weiss2022tunable,nolen2023arbitrarily}. In all of these demonstrated devices, heating is all that is required to produce the desired light as the relevant optical modes are excited thermally, thus providing an elegant source of MIR radiation.

Imparting tunability into such devices - which could allow for dynamic beam control and frequency shifting - requires the integration of materials with variable optical properties.  Materials with temperature-dependent phases and/or indices, such as GST\cite{chu2016active,abdollahramezani2022electrically,zhang2021electrically}, VO$\_{2}$\cite{rensberg2016active,nouman2018vanadium,kim2019phase,audhkhasi2021vanadium}, or Si\cite{sun2013large} have been utilized to create metasurfaces that control the magnitude and phase of scattered light in reflection or transmission geometries, but such materials are unsuitable for thermal emission devices that operate at high, constant temperatures. Alternatively, materials with indices that depend on carrier density, including graphene, III-V quantum wells and indium tin oxide (ITO), can be utilized to bestow electrostatic tunability on metasurfaces, and devices that control phase, frequency, and intensity of reflected light have been demonstrated\cite{Perfect_absorption_Kim,  Graphene_phase_modulation_Kim, 2D_MIR_review_Kim,huang2016gate,sarma2018metasurface,wu2019dynamic}. These materials are also chemically and phase stable at high temperatures, which has enabled them to be integrated within thermal-control metasurfaces to electrostatically tune the intensity and frequency of incandescent light in the mid-IR\cite{Graphene_thermal_modulation_Brar, Dynamic_thermal_emission_Brongersma,Dynamic_thermal_emission_Inoue}. Unfortunately, such materials also introduce ohmic loss which can, in some geometries, suppresses formation of the long-range delocalized modes that are necessary for coherent, directional thermal emission. As such, dynamic angular tuning of thermal emission is an outstanding problem in the field of thermal metasurfaces.

In this work, we theoretically describe and experimentally demonstrate a thermal emission device that can be tuned electrostatically to control the directionality of thermal emission within a narrow bandwidth.  We show experimentally that by using a tunable graphene-integrated metasurface as a boundary for a delocalized F-P cavity mode, the thermal emission from a surface at \SI{6.61}{\um} (\num{1508} \unit{cm\^{-1}}) can be continuously steered by $\pm$ \ang{16} by changing the carrier density of the graphene sheet.  Theoretical calculations, meanwhile, show that an optimized geometry using real materials could achieve $\pm$ \ang{60} of continuous tuning.

\section*{Results}

For dynamic thermal emission steering, we utilize an electrically tunable F-P resonance of a SiN$_\TU{x}$  dielectric layer sandwiched by a gold back reflector and a graphene-based active metasurface as illustrated in Fig. 1(a). The graphene metasurface consists of \SI{30}{nm} thick, \SI{1}{\um} wide gold strips spaced 40 nm apart on top of HfO$_{2}$ (5 nm)/graphene/Al$_{2}$O$_{3}$ (\SI{30}{nm}) trilayer, sitting on the 2 \unit{\um} think SiN$_\TU{x}$ membrane with the 100 nm gold back reflector that also serves as a back gate electrode. The gaps between the gold strips are filled with a bilayer of \SI{30}{nm} gold and 100 nm SiO$_\TU{x}$. The sub-wavelength period of the structure suppresses far-field diffraction except for the zeroth order.  We note that this structure does support metal-insulator-metal (MIM) surface plasmon modes\cite{dionne2006plasmon}, but for the device dimensions used in this work the MIM resonances occur at frequencies ($\sim$4500 \unit{cm\^{-1}}) much higher than the active frequency ($\sim$1500 \unit{cm\^{-1}}), and thus have little effect on the thermal steering properties of the device.

The working principle of our device is illustrated in Fig. 1(b). The graphene-based metasurface covering the top surface of the SiN$_\TU{x}$ membrane acts as a partially reflecting mirror to form a vertical F-P cavity. By applying an electrostatic potential ($V_G$) across the dielectric spacer, the Fermi level of graphene ($E_F$) is modulated and so are the complex reflection and transmission coefficients of the top graphene metasurface. Consequently, the condition for the resonance shifts, causing a shift in the peak emission angle ($\theta$) for a given frequency. These changes can be qualitatively understood by treating the top metasurface as a two-dimensional sheet with an effective surface admittance, which is justified since the metasurface thickness is about two orders of magnitude shorter than the wavelength of the free space light\cite{Salisbury_Jang, Perfect_absorption_Kim, Complete_Complex_Han}. In this model, the subwavelength metallic stips with narrow gaps make the overall optical response of the graphene metasurface to be highly capacitive (i.e. large imaginary impedance) at a low carrier concentration. As the conductivity of graphene raises with increasing $E_F$, the metasurface exhibits a reduced, but still high, capacitance and also acquires a larger conductance, changing the reflection/transmission characteristics. The quantitative surface admittance model for the graphene metasurface is discussed in detail in Supplementary Notes 1, 2, and 3.

Recognizing the emissivity $\epsilon (\omega, \theta) $ of a reciprocal object is equal to its absorptivity $\alpha(\omega, \theta)$\cite{Abs_emit_Kirchoff}, one can understand the mechanism of the directional shift in thermal emission more intuitively by analyzing the absorption process. Since the transmission channel is blocked by the back reflector, 
\begin{equation}
    \epsilon = \alpha  = 1 - |r_\TU{tot}|^2 = 1- |r_\TU{direct} + r_\TU{FP}|^2,
\end{equation}
where $r_\TU{tot}$ is the total reflection, which can be decomposed into the direct reflection from the top surface ($r_\TU{direct}$) and the resonant reflection due to the F-P interference formed by multiple reflections inside the dielectric spacer ($r_\TU{FP}$). The interplay between $r_\TU{FP}$ and $r_\TU{direct}$, both of which are dependent on $E_F$, determines the overall absorption (and thus the emission) of the device. The absorption peak occurs when $r_\TU{FP}$ and $r_\TU{direct}$ destructively interfere with each other by having similar amplitudes and a $\pi$ phase difference.

We first theoretically investigate the behavior of the proposed device using full-field electromagnetic simulations based on the finite element method as summarized in Fig. 2. The dependence of $r_\TU{direct}$ on $\theta$ and $E_F$ for TM polarized light is shown in Fig.2(b). $r_\TU{direct}$ can be obtained by simulating the reflection by the graphene metasurface sitting on a semi-infinite SiN$_\TU{x}$ layer without a back reflector. Since the top graphene metasurface does not support any distinctive resonance around the target frequency of $\omega = 1503$ \unit{cm^{-1}}, the direct reflectance, $R_\TU{direct}=|r_\TU{direct}|^2$, exhibit a generic weak dependence on $\theta$ within the range of \ang{0} to \ang{50}. As the carrier density of graphene increases, the metasurface becomes less capacitive, leading to better impedance matching as elaborated in (Supplementary Notes 1 and 3). Consequently, $R_\TU{direct}$ monotonically decreases with increasing $E_F$. The phase of the direct reflection, $\phi_\TU{direct} = \arg \{r_\TU{direct}\}$, remains nearly constant round 0.9$\pi$ within $\theta\in(\ang{0},\ang{50})$ and $E_F\in(0.2,0,65) \unit{eV}$.

Unlike $r_\TU{direct}$, $r_\TU{FP}$ shows a strong dependence on both $\theta$ and $E_F$ due to its resonant nature. The F-P resonance occurs when the out-of-plane wavevector inside the dielectric, $k_\TU{out} = nk_0 \sqrt{1 - \sin^2\theta}$, satisfies the constructive interference condition: 
\begin{equation}
    2k_\TU{out}h + \phi_\TU{top} + \phi_\TU{bottom} = 2{\pi} m,
\end{equation}
where $k_\TU{out}h$ is the phase accumulation associated with vertical wave propagation across the dielectric spacer, $\phi_\TU{top}$ and $\phi_\TU{bottom}$ are the reflection phase from the top and bottom surfaces, respectively, and $m$ is an integer. $\phi_\TU{bottom}\sim \pi$ does not dependent on $E_F$ since the bottom surface is a mere gold back reflector, which behaves like a perfect electric conductor at mid-infrared frequencies. $\phi_\TU{top}$, in principle, could depend on $E_F$ for metasurfaces with an admittance comparable to the surrounding medium, but in our device the admittance is large and, thus, the dependence of $\phi_\TU{top}$ is weak for $E_F \in (0.2, 0.65)$ \unit{eV}. (see Supplementary Note 1 for a detailed analysis). As a result, at a fixed frequency, the resonance angle $\theta_\TU{FP}$ slightly decreases from \ang{35} to \ang{29} when $E_F$ increases from \SI{0.2}{eV} to \SI{0.65}{eV} as indicated as a blue dashed curve in Fig.2(a); And, at a fixed $\theta$, the resonance frequency $\omega_\TU{FP}$ slightly blueshifts with increasing $E_F$. The F-P resonance becomes weaker with increasing $E_F$ as the top graphene metasurface becomes less reflective and more absorptive, raising both the radiative and dissipative decay rate of the resonant mode. However, while $\phi_\TU{top}$ shows only a small dependence on $E_F$, the overall phase shift due to the F-P resonance ($\phi_\TU{FP}$) includes phase accumulated while passing into and out of the F-P cavity, through the complex transmission coefficients $t_\TU{in}$ and $t_\TU{out}$, which show considerably more dependence on $E_F$. (see Supplementary Note 3)

Since the amplitude of $r_\TU{FP}$ is similar to that of $r_\TU{direct}$ near the broad F-P resonance, what mainly determines the overall absorption is their phase difference, $\phi_\TU{FP} - \phi_\TU{direct}$. We note that the Fano interference between a non-resonant and a resonant scattering channel has been widely adopted to create a sharp resonant response\cite{Fano_review_Kivshar, WG_modulator_Jang}. The dependence of $\phi_\TU{FP}-\phi_\TU{direct}$ on $E_F$ and $\theta$, which is dominated by $\phi_\TU{FP}$ due to the near constant $\phi_\TU{direct}\approx 0.9\pi$, are plotted in Fig. 2(c). $\phi_\TU{FP}$ monotonically decreases with $\theta$ because the propagation phase across the dielectric spacer, $k_\TU{out}h$, decreases as $k_\TU{out}$ shortens. $\phi_\TU{FP}$ also decreases with $E_F$ as the capacitive phase shift of the top graphene metasurface reduces. As a result, the condition for the Fano resonance, $\phi_\TU{FP}-\phi_\TU{direct}=\pi$, shifts from $\theta_\TU{res}=\ang{32}$ to \ang{0} as $E_F$ alters from \SI{0.2}{eV} to \SI{0.62}{eV}.  This change in the phase matching condition drives an overall change in the angular-dependent absorptivity/emissivity, shown in Fig.2(d), and thus allows the device to thermally emit at an angle that can be tuned by varying $E_F$. 

In order to experimentally verify the possibility of active thermal emission steering, we fabricated the proposed device using e-beam lithography over a $4\times4$ \unit{mm^2} area (see Methods), heated it to \SI{250}{\degreeCelsius}, and measured its angle-dependent thermal emission spectra while varying the $E_F$ by applying different gate voltages $V_G$. A polarizer was used to accept only TM polarized emission, and the acceptance angle of the emitted light was \ang{3}. The emissivity of the structure is calculated by normalizing the emitted radiation of the device to the emitted radiation of a reference carbon nanotube blackbody\cite{Carbon_nanotube_Mikhail}.

The measured surface normal emissivity spectra for $\theta=\ang{0}$ at $V_G$ = 560, 0 and \num{-560} V, shown in Fig.3(a), exhibit a well-defined resonance peak at around \num{1,500} \unit{cm^{-1}} that blueshifts as the Fermi level of graphene increases, indicating that the thermal emission peaks are electrostatically tunable with minor variation in the intensity. The measured emissivity spectra also shows a strong angular dependence as shown in Fig.3(b). At a constant doping level ($V_G=-560$ V), the emission peak shifts from \SI{1508}{ cm^{-1}} to \SI{1543}{cm^{-1}} as $\theta$ changes from \ang{0} to \ang{30}. There are also higher order features present around 2400 \unit{\cm^{-1}} (see Supplementary Note 4) that show similar but more limited shifting. Finally, Fig.3(c) demonstrates the dynamic thermal emission steering by showing how the emission angle is modulated by altering the doping level of graphene at a fixed target frequency $\omega = 1508$ \unit{cm^{-1}}. At $V_G=-560$ \unit{V}, we observe that the emission peak is most intense at normal incidence and decreases in intensity as the angle is increased. As the applied gate voltage increases to \SI{560}{V}, the lobe shifts from normal incidence to increasing angles, up to \ang{16}, allowing for continuous tuning in that range.

These experimental results can be compared to simulated emissivity spectra shown in Figures 3(d-f). In these simulations, the value of $E_F$ at $V_G= 0$ \unit{V}  was chosen as a fitting parameter and calculated spectra were compared to the experimental spectra obtained at $V_G= 0$ \unit{V} to determine that $E_F = -0.55$ \unit{eV} with no gate voltage applied.  This indicates that the sample is heavily hole-doped, which is consistent with previous studies of graphene grown and transferred using similar procedures\cite{Graphene_plasmon_Joel}. Using this initial value of $E_F$, the Fermi energies at other gate voltages were derived with a simple capacitance model. 

The overall qualitative behavior of the simulations (Figs. 3(d-f)) is consistent with our experimental results (Figs. 3(a-c)), however, the emission lobes are broader and the change of emission angle of the emitter is smaller in our experiment than was theoretically predicted. The likely sources of these inconsistencies are the metastructure geometric and material parameter variations across the full 4 x 4 mm$^2$ device (see Subsections A and B in Supplementary Note 5), and carrier density variation during the heating process due to the temperature dependence of the SiN$_\TU{x}$, Al$_{2}$O$_{3}$, and HfO$_{2}$  dielectric properties\cite{allers2004prediction,bellucci2021dielectric,bulbul2007frequency,chen2004temperature,dow2017thermal,zhu2002current}. The estimated carrier density tuning range is also affected by substrate and interface charge traps, which can act to decrease the overall doping range by screening the applied gating field(see Methods); Moreover, the tuning range is directly affected by the DC dielectric constant of the SiN$_\TU{x}$ layer, which has reported variations of 15$\%$ for the commercial membranes used in this work.\cite{norcada}  We also note that the modulation depth at $\theta \approx \ang{0}$ is predicted to be larger than what is observed experimentally, and we also attribute this mostly to decreases in the doping range, as well as small potential misalignment of the heating stage. (see Methods)  The intensity of emission at large angles can also be reduced due to ellipsoidal elongation of the measurement area which, for small device areas, can extend the active zone to include some low emissivity, unpatterned gold areas.   Finally, we note that the initial graphene doping level can affect the calculated spectra, but simulations show that such affects do not account for the discrepancies we observe (see Supplementary Note 5).  Those simulation results do indicate, however, that the dynamic steering range could be improved in the future by using graphene with slightly less initial doping.  

To further explore the potential performance of the proposed thermal steerer device, we investigate the maximum realizable emission angle under the limitation of realistic geometric and material parameters. The Fermi level of graphene is assumed to be electrostatically tunable between \SI{0}{eV} and \SI{0.6}{eV}, considering typical dielectric strength of SiN$_\TU{x}$ and numerical optimizations of the geometric parameters of the device were performed to maximize the angle tunability. To prevent performance degradation due to non-local effects (see Supplementary Note 5 for more discussion), we set the minimum gap width to \SI{30}{nm} and carried out simulations in the frame of classical electrodynamics. Figure 4(a) shows the structure of the optimized device. The gap and width of Au slit array are \SI{30}{nm} and \SI{740}{nm}, respectively. The HfO$_{2}$ is thinned to 1 nm which is achievable smallest value that could avoid quantum tunneling effect. The bilayer Au/SiO$_\TU{x}$ area eliminated to enhance interaction between graphene and Au slit array. The optimization results show that it is possible to achieve $\sim$ \ang{60} thermal emission angle steering with unity peak emissivity (Fig.4(b)).  The achievable performance is greater than most metasurface-based electrically tunable beam steering devices\cite{Beam_steering_Pierre} and is comparable to state-of-the-art MEMS-based beam steering device where field of view\cite{MEMS_LiDAR_Xiaosheng}.  The improvements in the optimized structure in comparison to the experimentally measured sample are due to three main effects. First, the optimized structure utilized a smaller, \SI{30}{nm} spacing between the gold strips.  This acts to increase the electric field concentration within the graphene and minimize stray fields connecting the gold strips, allowing more interaction with the graphene and a stronger effect of the graphene on the metasurface properties.  Second, a thinner HfO$_{2}$ layer is used in the optimized structure, which brings the graphene closer to the gold and also increases the electric field intensity within the graphene sheet (see Supplementary Notes 2 and 5). And, third, in the optimized structure we assume a greater range of $E_F$ tunability, which is consistent with the potential properties of the dielectrics, but could not be achieved in our experiments due to our methods of contacting the sample (i.e. wirebonding) which weakened the dielectric strength and restricted the range of $V_G$. We note that the required gate voltage for device operation can be significantly reduced by modifying the gating scheme. For example, as demonstrated in the work of N. H. Tu\cite{Graphene_ZnO_modulation_Tu} and B. Zeng\cite{High_speed_graphene_metasurface_Zeng} by inserting a transparent conducting layer near the top graphene membrane, the gate voltage required to achieve the same level of Fermi energy can be reduced by orders of magnitudes with a marginal perturbation of the optical characteristics of the device. 

\section*{Discussion}

In conclusion, we have demonstrated a thermal emitter that can continuously change the angle of emission in the mid-IR for a designated frequency. We show that by including a graphene-metal metasurface as a boundary, a delocalized F-P optical mode can be tuned to exhibit resonances with angular and frequency dependencies that depend on the carrier density of graphene, which can be tuned electrostatically. The net result is a surface that has an emissivity that is strongly angular dependent and tunable. 16$^{\circ}$ of thermal emission steering at \SI{6.61}{\um} was demonstrated experimentally, and we outline design strategies that could increase the tunability to almost 60$^{\circ}$. This work lays the foundation for next generation beamsteering devices that do not require an external lightsource, and could be broadly applicable for remote sensing and thermal camouflage applications.

\section*{Methods}

\subsection*{Fabrication of Device}
SiN$_\TU{x}$ membranes (\SI{2}{\um} thick and \SI{5}{mm} x \SI{5}{mm} wide) on a \SI{200}{\um} Si frame were purchased from Norcada. To the backside of the SiN$_\TU{x}$ membrane, we deposited a 2.5 nm chromium adhesion layer followed by a 100 nm of gold, which makes the lower layer opaque and reflective. Atomic Layer Deposition (a Fiji G2 ALD) was used to grow a \SI{30}{nm} film of Al$_2$O$_3$ on the top of the SiN$_\TU{x}$ membrane. Once the Al$_2$O$_3$ was grown, a prepared graphene sheet was transferred on top of the Al$_2$O$_3$ film. Graphene was purchased from Grolltex and was grown on a Cu foil. To remove the foil, first a protective layer of PMMA (950k, A4, MicroChem Corp.) was added on top of the graphene. The Cu foil was etched away with FeCl$_3$ (CE-100, Transene) then the graphene/PMMA stack was rinsed in a series of deionized water baths until transfer to the prepared membranes. Once transferred, the PMMA was removed by soaking in \SI{60}{\degreeCelsius} acetone for 1 h. After the graphene transfer, a \SI{5}{nm} film of HfO$_2$ was grown via atomic layer deposition. To prepare the SiN$_\TU{x}$ membranes for the next steps, the Si frame of the sample was glued to a carrier Si chip with PMMA (950k, A8, MicroChem Corp.). The prepared substrate was then coated with a negative tone hydrogen silesquioxane resist (HSiQ, 6\%, DisChem Inc.) at \SI{100}{nm}. The sample was then exposed and patterned using the Elionix ELS G-100, an electron beam lithography tool. After exposure, the samples were developed in MF-321 for 90 s, with a 30 s rinse in DI water and then a 30 s rinse in IPA.  The development process converts the exposed HSiQ to SiO$_\TU{x}$. For metal deposition of the top, a metal mask was placed above the substrate to create electrically disconnected regions. The deposition consisted of a \SI{2.5}{\nm} chromium adhesion layer and \SI{30}{\nm} of gold.  Following these processing steps, the graphene was found to be heavily hole-doped, similar to what has been observed in previous works\cite{Graphene_plasmon_Joel, Graphene_thermal_modulation_Brar},  Gate-dependent resistivity measurements showed an increase in resistance for positive gate bias, but no maximum resistance was observed that would indicate charge neutrality.  These measurements also exhibited hysteresis, consistent with what has been observed elsewhere, and indicative of surface, interface, and substrate charge traps that can be populated with charge as $V_G$ is changed.  At high biases, these traps can screen the applied gating field without doping the graphene, leading to deviations from the simple capacitance model that we use to estimate the graphene carrier density for a given $V_G$\cite{bonmann2017effects,lu2022hysteresis,bartosik2020mechanism}.

\subsection*{Thermal Emission Measurements}
The emission measurements were performed using a Bruker Vertex 70 FTIR, where thermal emission from a heated sample was used as the lightsource of the interferometer. The device was mounted on a rotation stage, and thermal emission from the device is collected by the FTIR\cite{Carbon_nanotube_Mikhail}. A carbon-nanontube source was used as our blackbody reference measurement. The finite size of the aperture creates a 3$^{\circ}$ acceptance angle, and there is also some uncertainty in the overall angle due to mechanical play in the stage holder and sample tilting within the sample holder. We estimate this uncertainly to be $\leq 3^{\circ}$ based on measurements with an alignment laser reflected off of an unpatterned area of the sample surface.

\subsection*{Optical Simulations}
The frequency-dependent dielectric functions of Al$_{2}$O$_{3}$, Cr, Au and SiO$_\TU{x}$ were taken from the Palik data\cite{Mat_data_Palik}. The dielectric functions of HfO$_{2}$ and SiN$_\TU{x}$ were obtained from infrared ellipsometry\cite{Perfect_absorption_Kim}. Heat-induced dielectric function change of SiN$_\TU{x}$ is corrected through the higher-order F-P resonance peak which is insensitive to Fermi level modulation (see Supplementary Note 4). The graphene was modeled as a layer with zero thickness, and its optical conductivity was calculated by Kubo formula \cite{Kubo_Falkovsky}. The carrier mobility of graphene is assumed to be \SI{300}{cm^2/Vs} which is comparable to a previously reported value \cite{Perfect_absorption_Kim}. The reflection/transmission coefficients and absorption spectrum of the proposed structure were calculated by full-wave simulation with the finite element method. We determined the Fermi level of graphene at $V_G$ = 0 V by comparing the calculated and measured frequency and angular emissivity spectra. This process aimed to minimize the difference in various parameters, including resonance frequency, intensity and full-width half maximum. Throughout the estimation process, we constrained the expected Fermi level to the range of \SI{-0.45}{} to \SI{-0.55}{eV}\cite{Graphene_plasmon_Joel}, consistent with the hole-doped state observed at the $V_G$ = 560 V.

\section*{Data Availability}
The experimental and theoretical data used to generate the figures in this manuscript and in the Supporting Information are available on Zenado [10.5281/zenodo.10615359].

\def\bibsection{\section*{\refname}} 
\bibliographystyle{naturemag}
\bibliography{Main}

\section*{Acknowledgements}
J.S. and V.W.B. were supported by the Gordon and Betty Moore Foundation through a Moore Inventors Fellowship. M.F. was supported by Office of Naval Research award N00014-20-1-2356. M.A.K. and C.W. acknowledge support from the US Office of Naval Research, award N00014-20-1-2297. This work was also supported by the National Research Foundation of Korea (NRF) grants funded by the Ministry of Science and ICT (NRF-2022R1A2C2092095, S.K. and M.S.J.) and by the Ministry of Education (NRF-2022R1I1A1A01065727, S.K.).

\section*{Author Contributions}
S.K., M.S.J., and V.W.B. conceived the idea. J.S. fabricated and characterized the devices, and performed the optical measurements. M.F. assisted in sample fabrication.  M.A.K. and C.W. assisted in optical measurements. S.K. and M.S.J. conducted a theoretical analysis. S.K. conducted optical simulations and device optimization. L.S. and P.W. C.W. assisted in theoretical design and in fabrication planning. J.S., S.K., M.S.J., and V.W.B. wrote the manuscript. M.S.J. and V.W.B. supervised the project.

\section*{Competing interests}
The Authors declare no competing interests.
\clearpage

\newpage
\begin{figure*}[ht]
\includegraphics[width = 1\linewidth]{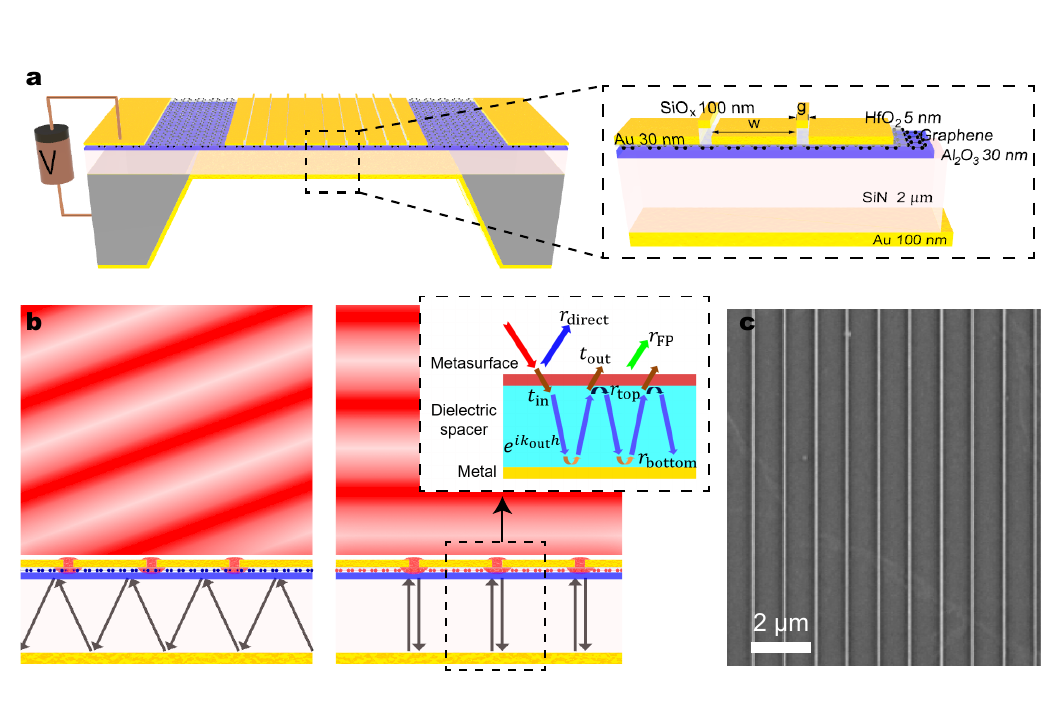}
\caption*{\textbf{Fig. 1. Schematic and working principles of dynamic thermal steering device.} \textbf{a} Diagram of the thermal steering device. The magnified view shows the geometry of a graphene-Au slit metasurface unit cell. \textbf{b} Illustration of the working mechanism of electrically tunable directional thermal emission via graphene metasurface control of the delocalized Fabry-Perot modes in the dielectric. The emission angle of the structure is controlled by the incident angle-dependent resonant absorption condition, which changes with the graphene Fermi level. The inset shows decomposed total reflection into two reflection channels: direct reflection $r_\TU{direct}$ and F-P reflection $r_\TU{FP}$.
\textbf{c} scanning electron microscopy image of graphene metasurface on top of SiN$_\TU{x}$ membrane. The width (w) and gap (g) of Au slit array are 1 \unit{\um} and 40 nm, respectively.
\vspace{-5mm}}
\end{figure*}

\begin{figure*}[ht]
\includegraphics[width = 1\linewidth]{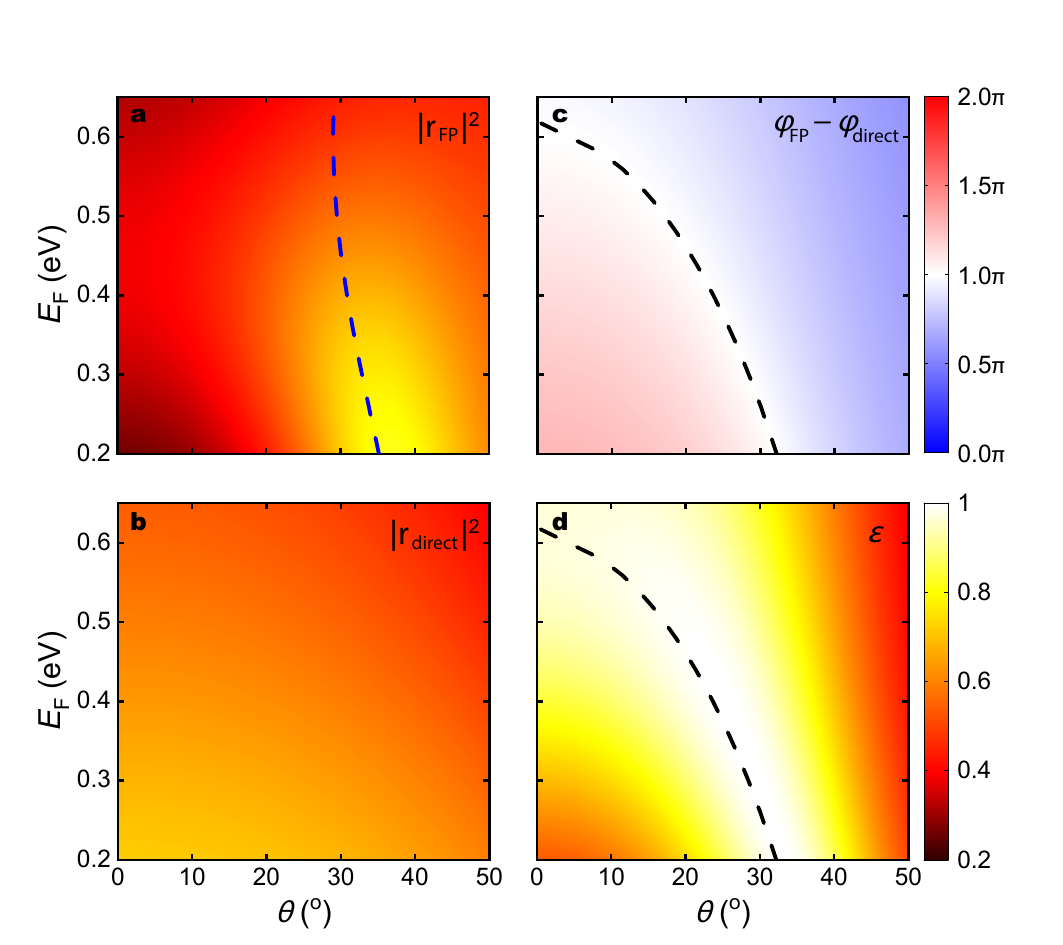}
\caption*{\textbf{Fig. 2. Angular and Fermi-level dependence of reflection coefficients and emissivity.} Fermi-level and angular dependence of \textbf{a} the reflectance due to the F-B resonance ($|r_\TU{FP}|^2$), \textbf{b} the direct reflectance from the top surface of the device ($|r_\TU{direct}|^2$), \textbf{c} The phase difference between the two reflection ($\phi_\TU{FP} - \phi_\TU{direct}$), and \textbf{d} the emissivity of the device ($\epsilon$). The blue dashed line in (\textbf{a}) indicates the F-P resonance condition. The black dashed line in (\textbf{c}) and (\textbf{d}) indicates the condition for destructive interference between $r_\TU{FP}$ and $r_\TU{direct}$, $|\phi_\TU{FP} - \phi_\TU{direct}|  = \pi$. All angular spectra are calculated for frequency $\omega = 1503$ \unit{cm^{-1}}.    
\vspace{-2mm}}
\end{figure*}

\begin{figure*}[ht]
\includegraphics[width = 1\linewidth]{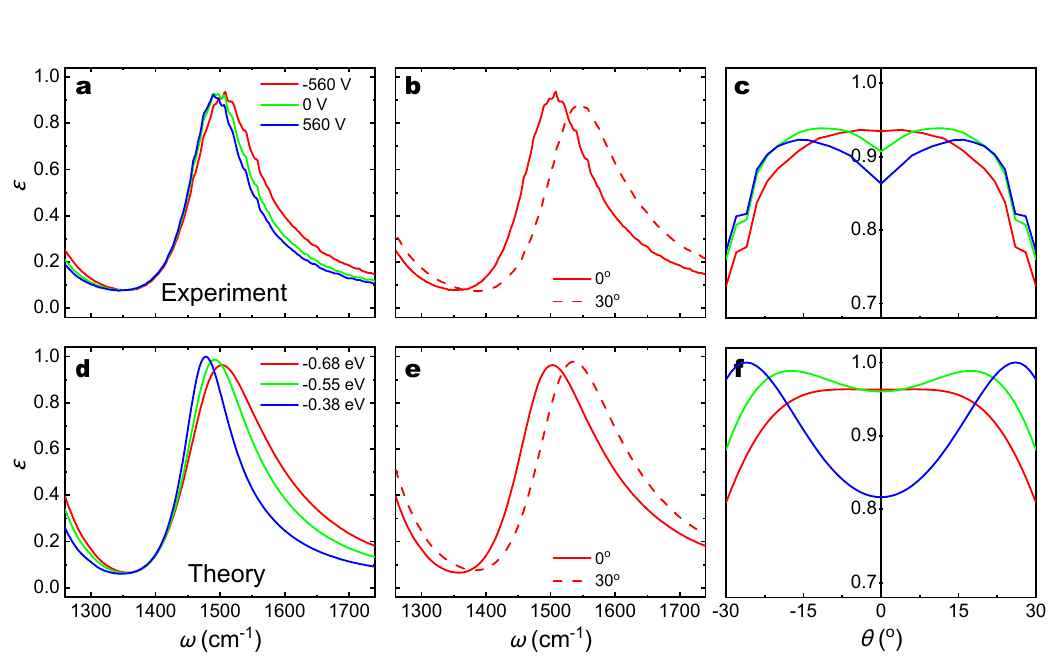}
\caption*{\textbf{Fig. 3. Measured and calculated emissivity dependence on gate voltage and measurement angle.} \textbf{a,b,c} Experimental emission spectra for (\textbf{a}) different applied gate voltages at normal incidence, (\textbf{b}) at different angles at a constant applied voltage of \SI{-560}{V}, and (\textbf{c}) the angular-dependent emission at \num{1508} \unit{cm^{-1}}. Experimental measurements were obtained for $\ang{0} < \theta < \ang{30}$ and are mirrored for visual clarity. \textbf{d,e,f} Show analogous simulated spectra, with \textbf{f} plotting  angular-dependent emission at \num{1503} \unit{cm^{-1}}.

\vspace{-2mm}}
\label{fig:fig_3}
\end{figure*}

\begin{figure}[ht]
\includegraphics[width = 1\linewidth]{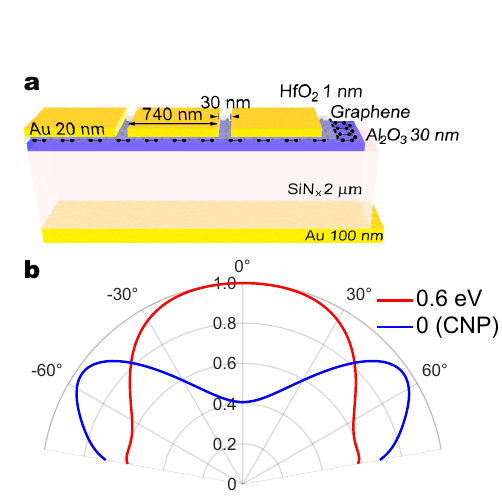}
\caption*{\textbf{Fig. 4. Optimized steering of thermal emission.} \textbf{a} Schematic of the device geometry that maximizes thermal emission steering performance. \textbf{b} The angular emission spectra of the optimized device for Fermi levels of \num{0} and \SI{0.6}{eV} at $\omega = 1614$\unit{cm^{-1}}.
\vspace{-2mm}}
\end{figure}

\end{document}

\maketitle

\pagebreak

\section*{Supplementary Note 1. Derivation of surface admittance of a metasurface}

\begin{figure*}[h]
\includegraphics[width = 0.5\linewidth]{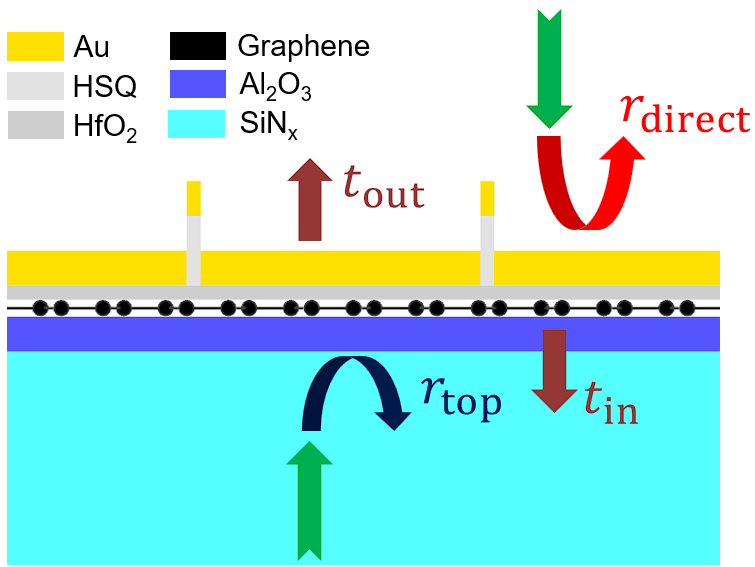}
\centering
\caption*{\textbf{Fig. S1.} Detailed configuration of metasurface and reflection/transmission coefficients. 
\vspace{-2mm}}
\end{figure*}

The metasurface can be modeled as a thin conductive layer with effective surface admittance $Y_\TU{s}$ because the thickness of the metasurface is much thinner than the wavelength of the incident light\cite{Salisbury_Jang, Perfect_absorption_Kim}. For normally incident light, $Y_\TU{s}$ can be solved for from the boundary condition $H_\textup{z}(0^+) - H_\textup{z}(0^-) = Y_\textup{s}E_\textup{x}(0)$. For the metasurface on top of the semi-infinite dielectric layer (Fig. S1), the normalized surface admittance $\tilde{Y}_\TU{s}$ is expressed in terms of transmission and reflection coefficients ($t$ and $r$) of a metasurface as:
\vspace{5mm}
\begin{equation}
    \tilde{Y}_\textup{s,t} = \frac{Y_\TU{s,t}}{Y_0} = \frac{Y_\TU{i}}{Y_0}(\frac{2}{t} - 1 - \frac{Y_\TU{t}}{Y_\TU{i}})
\end{equation}
\begin{equation}
    \tilde{Y}_\TU{s,r} = \frac{Y_\TU{s,r}}{Y_0} = \frac{Y_\TU{i}}{Y_0}(\frac{1 -r}{1 + r} -  - \frac{Y_\TU{t}}{Y_\TU{i}})
    \vspace{5mm}
\end{equation}
Here, $Y_{0}$, $Y_\textup{i}$ and $Y_\textup{t}$ are the admittance of free space, input, and output media. For the top (bottom) excitation, $Y_\textup{i}$ and $Y_\textup{t}$ are air (SiN$_\TU{x}$) and SiN$_\TU{x}$ (air), respectively. Figure S2(b) shows that the values of calculated admittance $\tilde{Y}_\textup{s,t}$ and $\tilde{Y}_\textup{s,r}$ are different. The surface admittance for both directions would be identical if the metasurface had zero thickness due to the same tangential electric fields at the top and bottom surfaces. However, the finite thickness of metasurface invalidates that condition and leads to two different surface admittance. In Note 1, we use normalized surface admittance derived from transmission coefficients.

We note that the dynamic behavior of the metasurface due to the Fermi level of graphene could be understood from the analysis of the surface admittance components. The real part (normalized surface conductance $\tilde{G}_\textup{s}$) and imaginary part (normalized surface susceptance $\tilde{B}_\textup{s}$) of the normalized surface admittance $(\tilde{Y}_\textup{s} = \tilde{G}_\textup{s} - i\tilde{B}_{s})$ provide information on absorption and scattering of the metasurface\cite{Admittance_model}. The normalized surface conductance $\tilde{G}_\textup{s}$ indicates the strength of the scattering process with the absorption of the metasurface. Because metal has much higher conductivity than graphene, the conductance of the metasurface is approximated to the conductance of graphene. Figure S2(c) shows higher conductance of the metasurface is obtained with an increase (decrease) of the Fermi level (frequency). In contrast, the normalized surface susceptance $\tilde{B}_\textup{s}$ indicates the strength of the scattering process without absorption of the metasurface. The susceptance of the metasurface is determined by a capacitance ($C$) and an inductance ($L$) derived from geometry and material parameters, where the surface susceptance $B_{s}$ of metasurface is proportional to $( -i\omega L + (-i\omega C)^{-1} )^{-1}$ with the convention of $e^{-i\omega t}$. Since the period of the metal slit array is much shorter than a free space wavelength, the surface susceptance of the metasurface shows a capacitive response. We note that the kinetic inductance of graphene is inversely proportional to the optical conductivity of graphene\cite{Kinetic_inductance}. Thus higher Fermi level of graphene results in lowering surface susceptance of metasurface. 

\begin{figure*}[t]
\includegraphics[width = 1.0\linewidth]{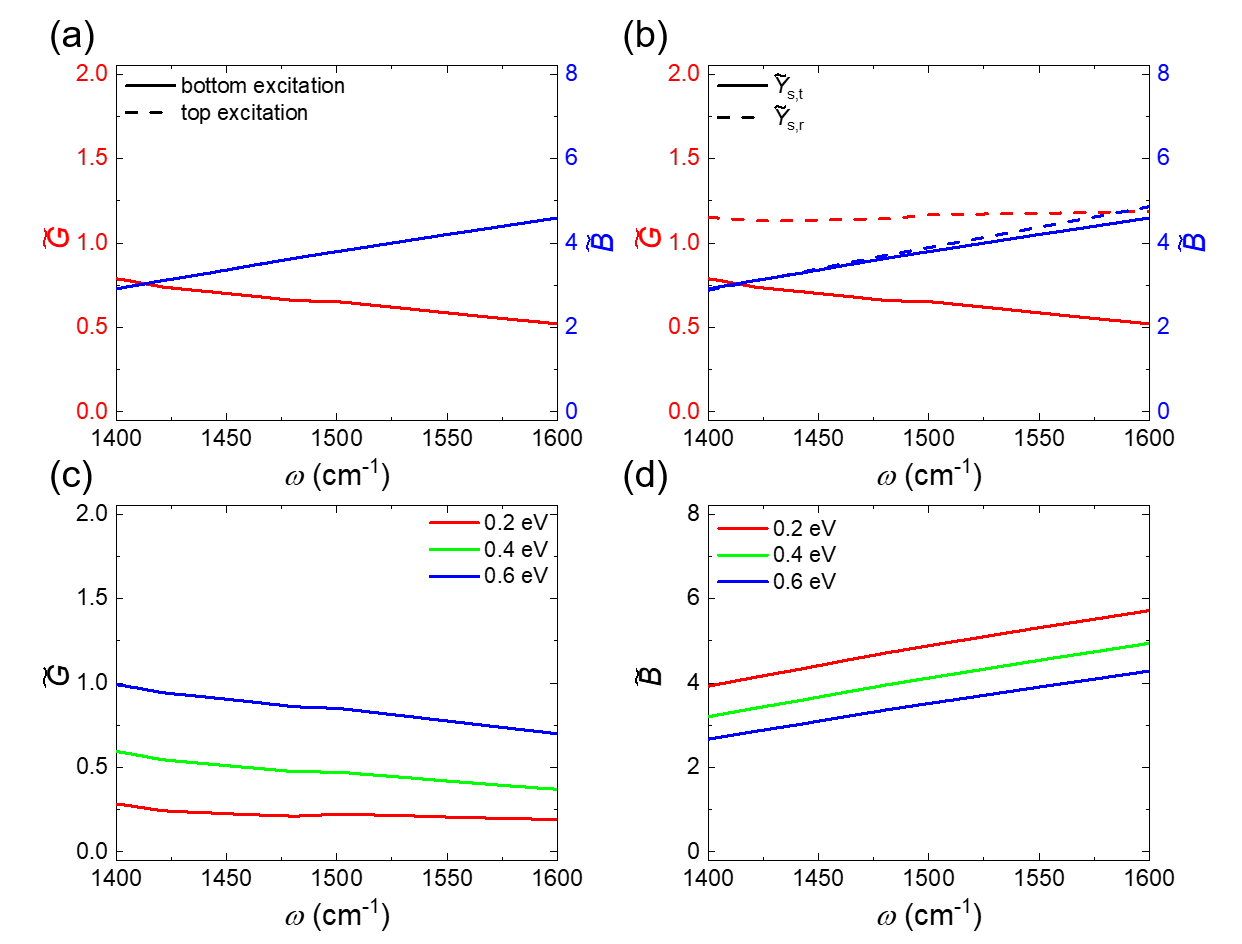}
\centering
\caption*{\textbf{Fig. S2. Analysis of metasurface admittance.} Conductance $\tilde{G}$ (left axis) and susceptance $\tilde{B}$ (right axis) of metasurface for different (a) excitation directions and (b) derivation variables ($r$ and $t$) for $\EF$ = \SI{0.5}{eV}. The (c) conductance $\tilde{G}$ and (d) susceptance $\tilde{B}$ for different Fermi levels of graphene.
\vspace{-2mm}}
\end{figure*}

\pagebreak

Next, we calculate angle-dependent normalized surface admittance of the surface current which could be expressed in terms of angle-dependent transmission and reflection coefficients as:

\begin{equation}
    \frac{Y_\TU{s,t}}{Y_\TU{i}} = \TU{sec} {\theta}_\TU{t} (\frac{2}{t} - \frac{\TU{cos} {\theta}_\TU{t}}{\TU{cos} {\theta}_\TU{i}} - \frac{Y_\TU{t}}{Y_\TU{i}})
\end{equation}

\begin{equation}
    \frac{Y_\TU{s,r}}{Y_\TU{i}} = \TU{sec} {\theta}_\TU{t} (\frac{\TU{cos} {\theta}_\TU{t}}{\TU{cos} {\theta}_\TU{i}} \frac{1 -r}{1 + r} -  - \frac{Y_\TU{t}}{Y_\TU{i}})
    \vspace{5mm}
\end{equation}
where the angles ${\theta}_\TU{i}$ and ${\theta}_\TU{t}$ represent the angles of incidence and transmission, respectively. Figure S3 shows that the surface admittance remains constant regardless of the incident angle which is consistent with the ideal surface current model. This implies that the surface admittance derived from normal incident light could be exploited to derive angle-dependent reflection and transmission coefficients of the metasurface.

\begin{figure*}[t]
\includegraphics[width = 1.0\linewidth]{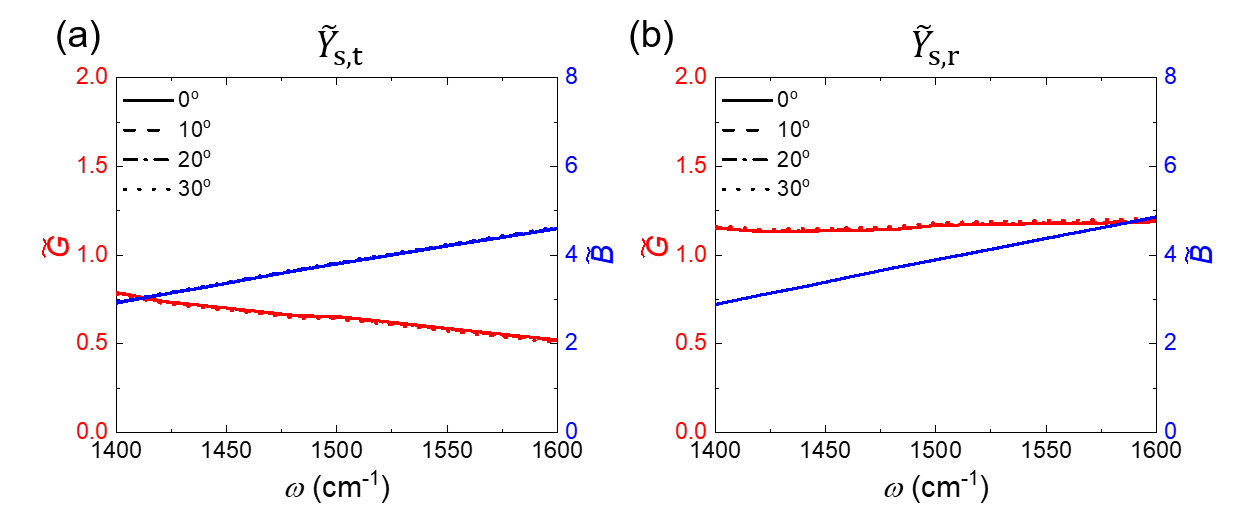}
\centering
\caption*{\textbf{Fig. S3. Dependence of metasurface admittance on illumination directions.} Conductance $\tilde{G}$ (left axis) and susceptance $\tilde{B}$ (right axis) of metasurface for different excitation directions derived from (a) transmission and (b) reflection coefficient. 
\vspace{-2mm}}
\end{figure*}

\pagebreak

\section*{Supplementary Note 2. Analysis of plasmonic structure in metasurface}

\begin{figure*}[h]
\includegraphics[width = 1.0\linewidth]{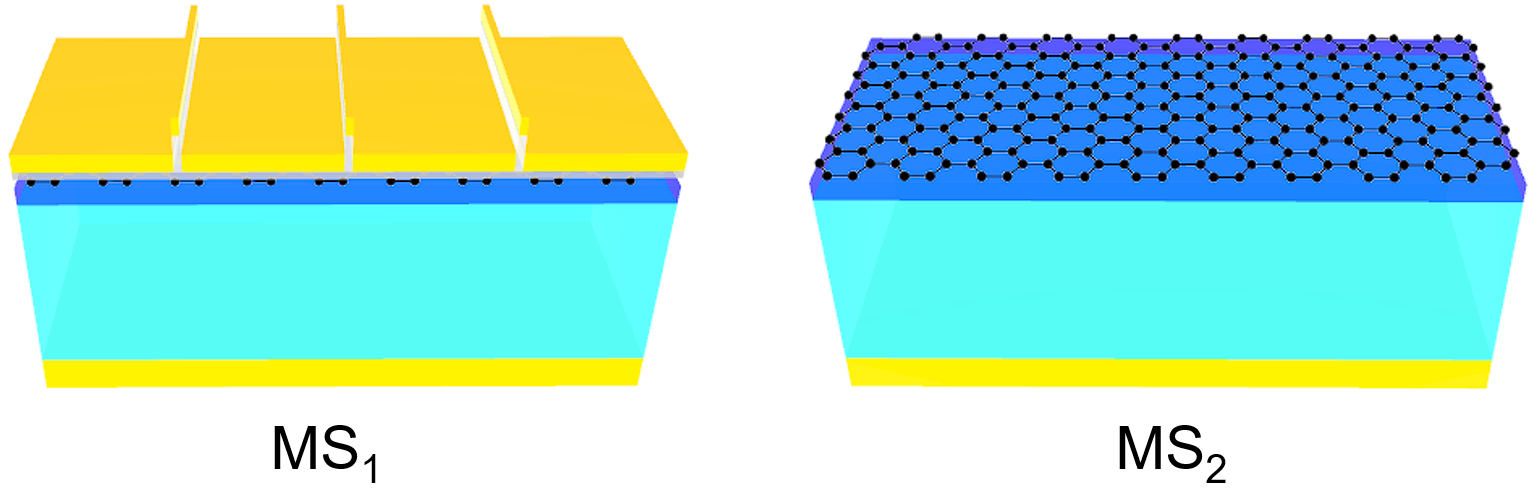}
\centering
\caption*{\textbf{Fig. S4.} Schematic of two different types of metasurfaces, MS$_{1}$ and MS$_{2}$, which differ based on the presence of a plasmonic metal slit array.
\vspace{-2mm}}
\end{figure*}

To investigate the impact of a plasmonic structure in the proposed graphene-metal hybrid metasurface, which serves as an electrically tunable mirror layer, we calculate the reflection coefficient for different geometric parameters of the plasmonic structure. The influence of the plasmonic structure is studied by comparing two different metasurface configurations MS$_{1}$ and MS$_{2}$ as shown in Fig. S4. Figure S5(a) and (b) illustrate the calculated amplitude ($|r_\TU{top, \SI{0.3}{eV}}|$) and phase difference ($\Delta\phi_\TU{top} = \phi_\TU{top,\SI{0.3}{eV}} - \phi_\TU{top,0 eV}$) of reflection coefficients as a function of gap width $g$ with Fermi levels \SI{0}{eV} and \SI{0.3}{eV}. By comparing Fig. S5(a) and (c), we observe that the amplitude of the reflection coefficient is mainly influenced by the slit width $w$, which represents the surface coverage of the highly reflective metal film. On the other hand, the phase modulation for the Fermi level is determined by the electric field intensity at the surface of graphene, which is determined by the compressed transmitted electromagnetic wave as the form of plasmonic wave. As a result, the phase difference is significantly affected by both geometric parameters. We emphasize that the shape of the resonance peak in the Fabry-Perot (F-P) resonator and the resonance frequency shift for the Fermi level modulation are determined by the reflection coefficient of the metasurface. Therefore, proper metasurface design is crucial for achieving a sufficiently high and sharp emissivity peak with a considerable resonance frequency shift for Fermi level modulation.

\begin{figure*}[p]
\includegraphics[width = 1.0\linewidth]{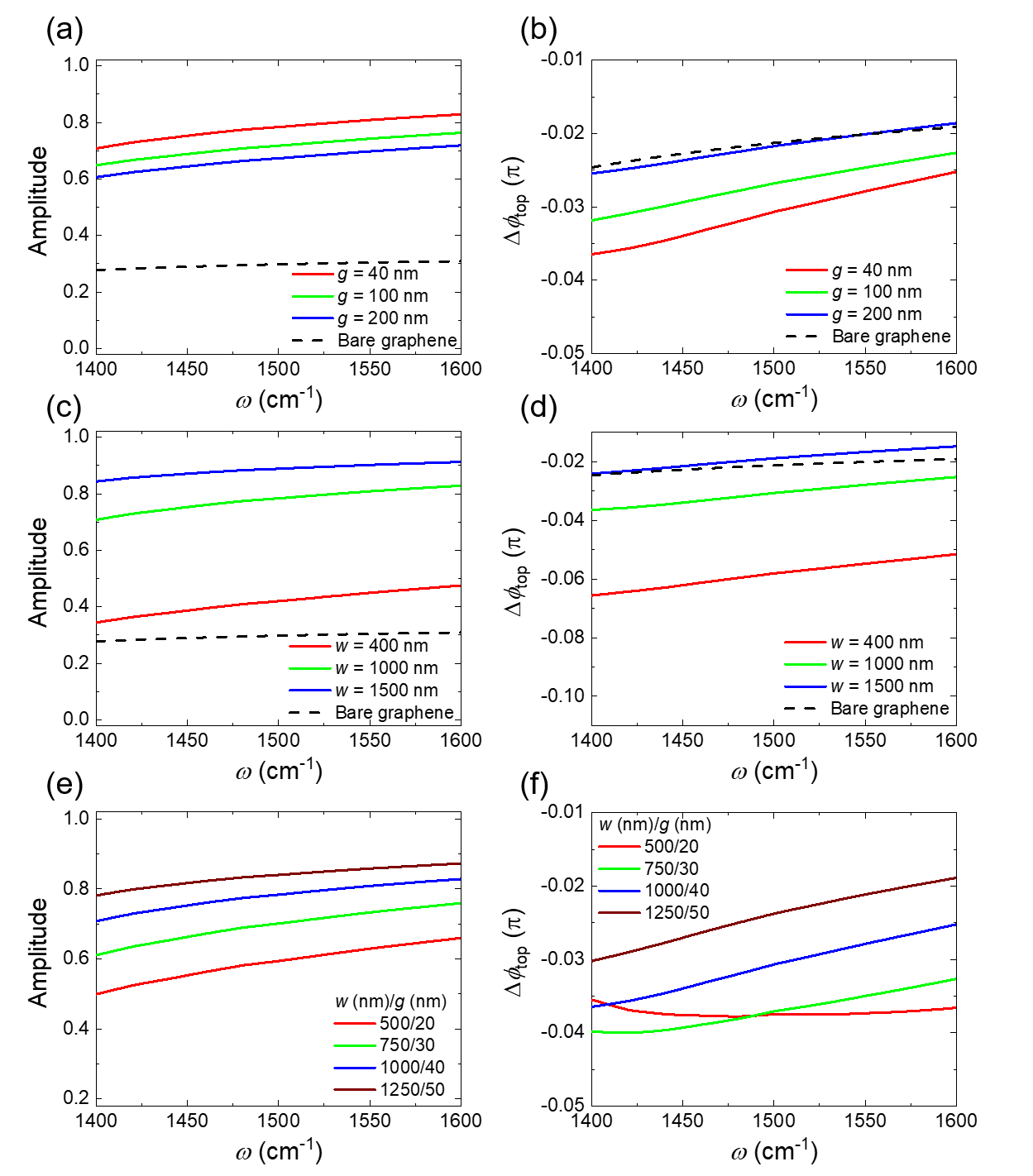}
\centering
\caption*{\textbf{Fig. S5. Influence of plasmonic metasurfaces on reflection coefficient modulation.} The amplitude $|r_\TU{top,\SI{0.3}{eV}}|$ and phase difference $\Delta\phi_\TU{top} = \phi_\TU{0.3 eV} - \phi_\TU{0 eV}$ for slit (a, b) and gap (c, d) widths g and w, respectively. The black dashed line shows the reflection coefficient of bare graphene structure MS$_{2}$. The amplitude (e) and phase difference (f) of metasurface MS$_{1}$ for constant ratio ($g/w) = 25$.
\vspace{-2mm}}
\end{figure*}

\pagebreak

\section*{Supplementary Note 3. Fabry-Perot model analysis}

\begin{figure*}[h]
\includegraphics[width = 0.5\linewidth]{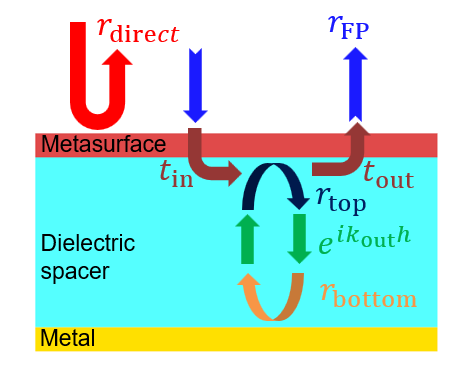}
\centering
\caption*{\textbf{Fig. S6.} Schematic of F-P model for a structure consisting of graphene-plasmonic metasurface/thick dielectric spacer/metal reflector.
\vspace{-2mm}}
\end{figure*}

In order to deeply understand the operation mechanism of the proposed structure, we develop the semi-analytical solution of the reflection coefficient for the proposed structure based on F-P interference (Fig. S6). To simplify the reflection coefficient equation of the structure, the dielectric film stack is merged into a single film, where reflection at the bottom interfaces of Al$_{2}$O$_{3}$ and HfO$_{2}$ films are ignored. The graphene layer thickness is excluded from the thickness of the dielectric stack because it was modeled as a zero-thickness conductivity sheet layer. Note that the subwavelength period and the non-resonant response of the metasurface suppress higher-order diffraction and deflection for incident light. We directly calculate the reflection ($r$) and transmission coefficients ($t$) from the definition in electromagnetic wave theory. The F-P reflection coefficient of the proposed structure is 
\vspace{5mm}

\begin{equation}
\rFP=\frac{t_\TU{in}t_\TU{out}r_\TU{bottom}e^{2{ik_{\TU{out}}}h}}{1-r_\TU{top}r_\TU{bottom} e^{2i{k_{\TU{out}}}h}}
\vspace{5mm}
\end{equation}
where $k_{\textup{out}}$ and $h$ are the out-of-plane wavevector and the thickness of the dielectric layer. The wavevector $k_{\textup{out}}$ is a function of refractive index and incident angle, and thus total phase accumulation is sum of $k_{\textup{out}}h$ at each film.

\vspace{2mm}

We note that a rich free electron density of noble metal makes it difficult to obtain a noticeable electro-optic effect in the bottom and top metal layers. Therefore, the modulation of the resonant frequency of the proposed structure is determined by the Fermi-level dependent surface admittance of the metasurface. Figure S7 shows the reflection and transmission coefficients for normally incident TM polarized plane wave as a function of Fermi levels. In the extreme case ($\EF \rightarrow $ \SI{0}{eV}), the amplitude and phase of $r_\TU{top}$ has high value of 0.82 and 0.83$\pi$ which is close to planar metal film because metal slit array covers $>95\%$ of surface area. In contrast, the increase (decrease) of conductance (susceptance) of the metasurface reduces the amplitude and phase of reflection coefficients for higher $\EF$ or lower $\omega$. Similarly, the transmission coefficients $t_{\TU{in},\TU{out}}$ also transit from a metal mirror-like response to a lossy dielectric response for $\EF$ and $\omega$.

\begin{figure*}[t]
\includegraphics[width = 1.0\linewidth]{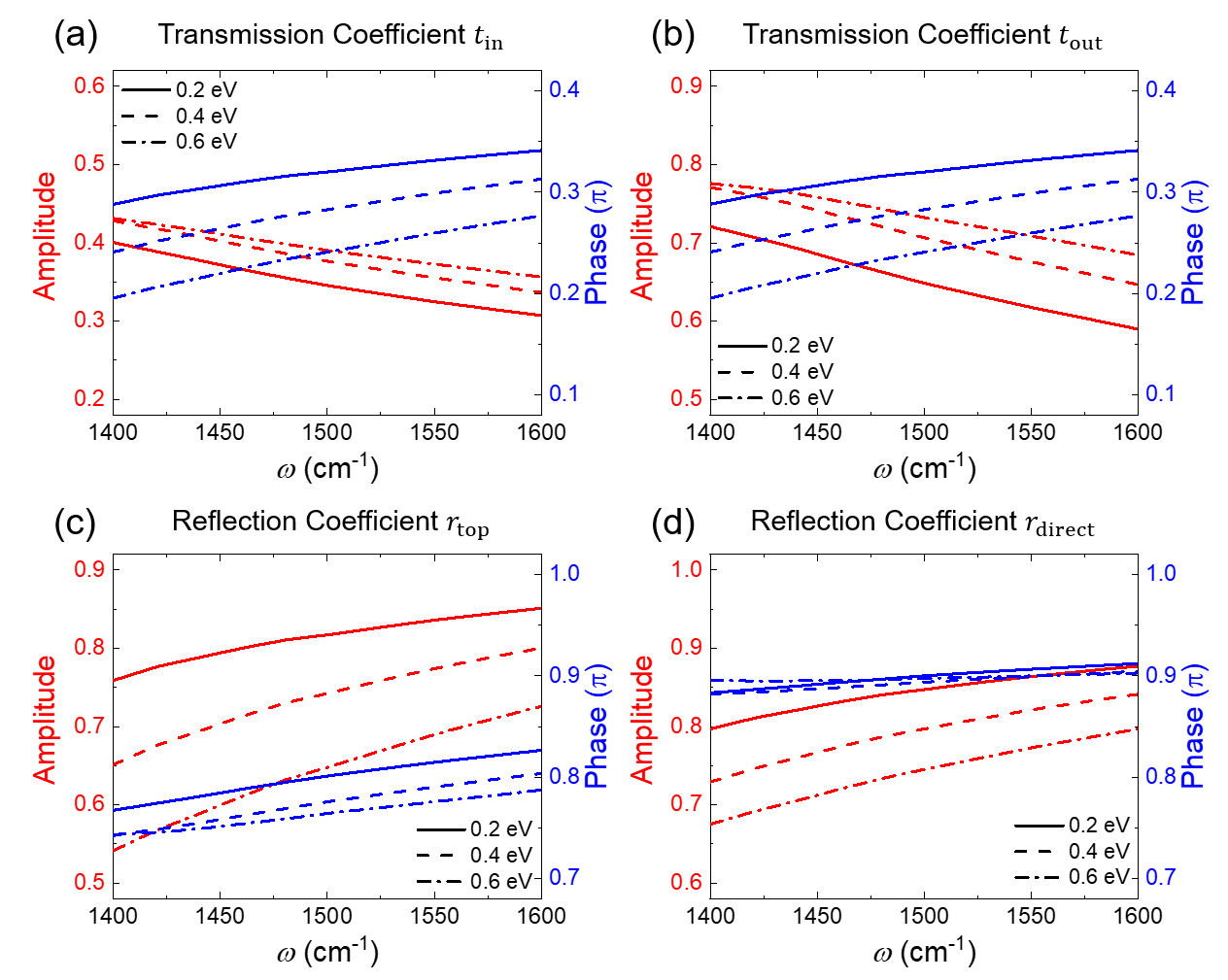}
\centering
\caption*{\textbf{Fig. S7. Fermi level dependence of transmission/reflection coefficients.} The amplitude (left axis) and phase (right axis) of transmission (a,b) and reflection (c,d) coefficients for different Fermi levels.
\vspace{-2mm}}
\end{figure*}

\begin{figure*}[hp]
\includegraphics[width = 1.0\linewidth]{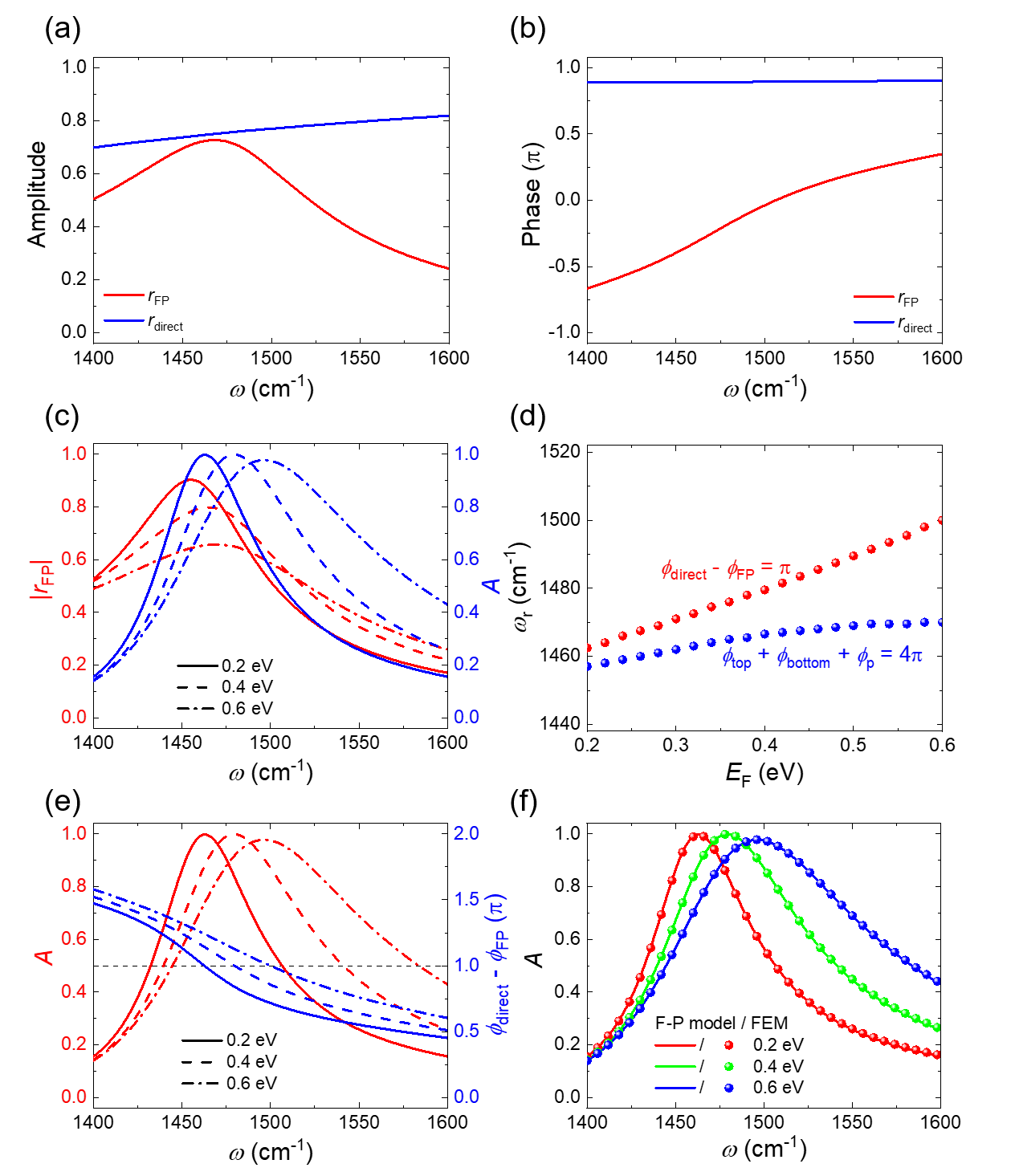}
\centering
\caption*{\textbf{Fig. S8. Analysis of Fermi level dependence using the semi-analytical model.} The (a) amplitude and (b) phase of reflection coefficient $\rFP$ and $\rd$ for Fermi level $E_F$ =  0.5 eV (c) The amplitude of $\rFP$ (left axis) and total absorption derived from F-P model (right axis). (d) The resonance frequencies $\omr$ and $\omFP$ as a function of graphene Fermi levels. (e) The F-P model total absorption (left axis) and phase difference (right axis) for different Fermi levels of graphene. (f) The total absorption from FEM-based full wave simulation and F-P model.
\vspace{-2mm}}
\end{figure*}

In the F-P model, the total absorption $A$ can be calculated as $1 - |r_\TU{tot}|^2$ = $1 - |\rd + \rFP|^2$. Note that the amplitude and phase variation ($\rd$/$\pd$) of the direct reflection is significantly smaller than the amplitude and phase ($\rFP$/$\pFP$) of the Fabry-Perot reflection as shown in Fig. S8(a) and (b). Thus, the dynamic behavior of total absorption is primarily determined by $\rFP$/$\pFP$. The dynamic behavior of these coefficients can be understood by analyzing the dependence of F-P reflection spectra $\rFP$ on the Fermi level and the incident angle. The resonance frequency of $\rFP$ is determined by the phase condition $2k_\textup{out}h + \phi_\TU{top} + \phi_\TU{bottom} = 2{\pi} m$, where $\phi_\TU{top}$ and $\phi_\TU{bottom}$ are the phase of reflection coefficient $r_\TU{top}$ and $r_\TU{bottom}$ respectively, and $m$ is an integer. Given the nearly constant reflection phase of the bottom electrode, the reflection phase change of metasurface $\Delta\phi_\TU{top}$ is compensated by the change of out-of-plane wavevector $k_\TU{out}$, which contributes to propagation phase $\phi_\TU{p} = 2\Delta k_\TU{out}h$. As $\phi_\TU{top}$ is inversely proportional to the Fermi level, the $k_\TU{out}$ at resonance frequency should be increased for higher Fermi level, leading to a blue shift in F-P resonance. Figure S8(d) shows that the resonance frequency $\omr$ derived from the phase condition for maximum absorption ($\pd - \pFP = \pi$) is larger than the F-P resonance frequency $\omFP$, and that the change of $\omr$ is faster than $\omFP$ for Fermi level modulation. At the F-P resonance condition, the phase of $\rFP$ is equal to the sum of phases of transmission coefficients $t_\TU{in}$ and $t_\TU{out}$ where the $\pi$ phase difference condition is not satisfied. Since $\pd - \pFP > \pi$ and $\partial(\pd - \pFP)/\partial\omega < 0$, the resonance frequency of maximum absorption $\omr$ becomes greater than F-P resonance frequency $\omr$. In addition, phase modulation of $\rd$, $t_\TU{in}$, and $t_\TU{out}$ for Fermi level provides additional phase difference between $\pd$ and $\pFP$, thus the resonance frequency shift of $\Delta\omr$ is larger than $\Delta\omFP$.

\begin{figure*}[t]
\includegraphics[width = 1.0\linewidth]{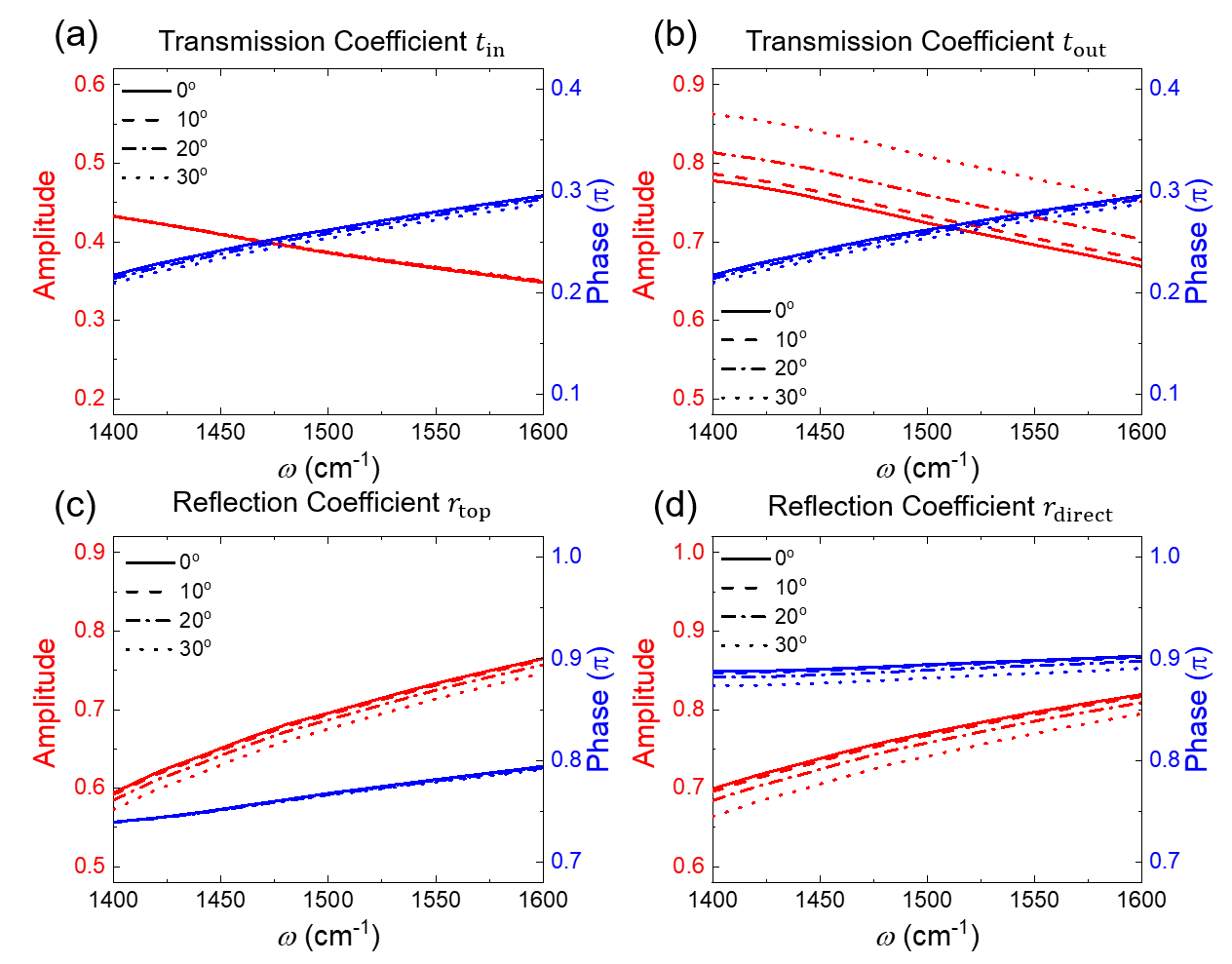}
\centering
\caption*{\textbf{Fig. S9. Incident angle dependence of transmission/reflection coefficients.} The amplitude (left axis) and phase (right axis) of transmission (a,b) and reflection (c,d) coefficients for different incident angles.
\vspace{-2mm}}
\end{figure*}

On the other hand, the parameters and variables in the F-P reflection formula also depend on the incident angle $\theta$ of the excitation light. Figure S9 shows the incident angle-dependent reflection and transmission coefficients. In the discussion of angle-dependent calculation, the incident angle is defined as the angle in the air. The amplitude and phase of coefficients show slower variation for incident angles than the Fermi level of graphene. Larger amplitude change of $t_\TU{out}$ than $t_\TU{in}$ originates from the input admittance difference (Eq. S1 and 2). We emphasize that the shift of F-P resonance frequency comes from phase accumulation change of propagating wave ($2\Delta k_\TU{out}h$) because the change of reflection and transmission coefficients for incident angles are smaller than the change of out-of-plane wavevector. The out-of-plane wavevector is given as $k_\TU{d}\TU{cos}\theta_\TU{d} = k_\TU{out}$ where $k_\TU{d}$ and $\theta_\TU{d}$ are the wavevector and the propagation angle in the dielectric spacer. Since the propagation angle has the relation with incident angle $\theta$ as $n_\TU{d}\TU{sin}\theta_\TU{d} = \TU{sin}{\theta}$, a larger incident angle reduces propagation phase accumulation. To compensate for the phase decrease by $\TU{cos}\theta_\TU{d}$, $k_\TU{d}$ should be increased, which is equivalent to a blue shift of resonance frequency. Unlike the resonance frequency change for Fermi level modulation, the resonance frequency change of $\Delta\omr$ and $\Delta\omFP$ for incident angle are similar due to the difference in the phase modulation method as shown in Fig. S10(a).

\begin{figure*}[t]
\includegraphics[width = 1.0\linewidth]{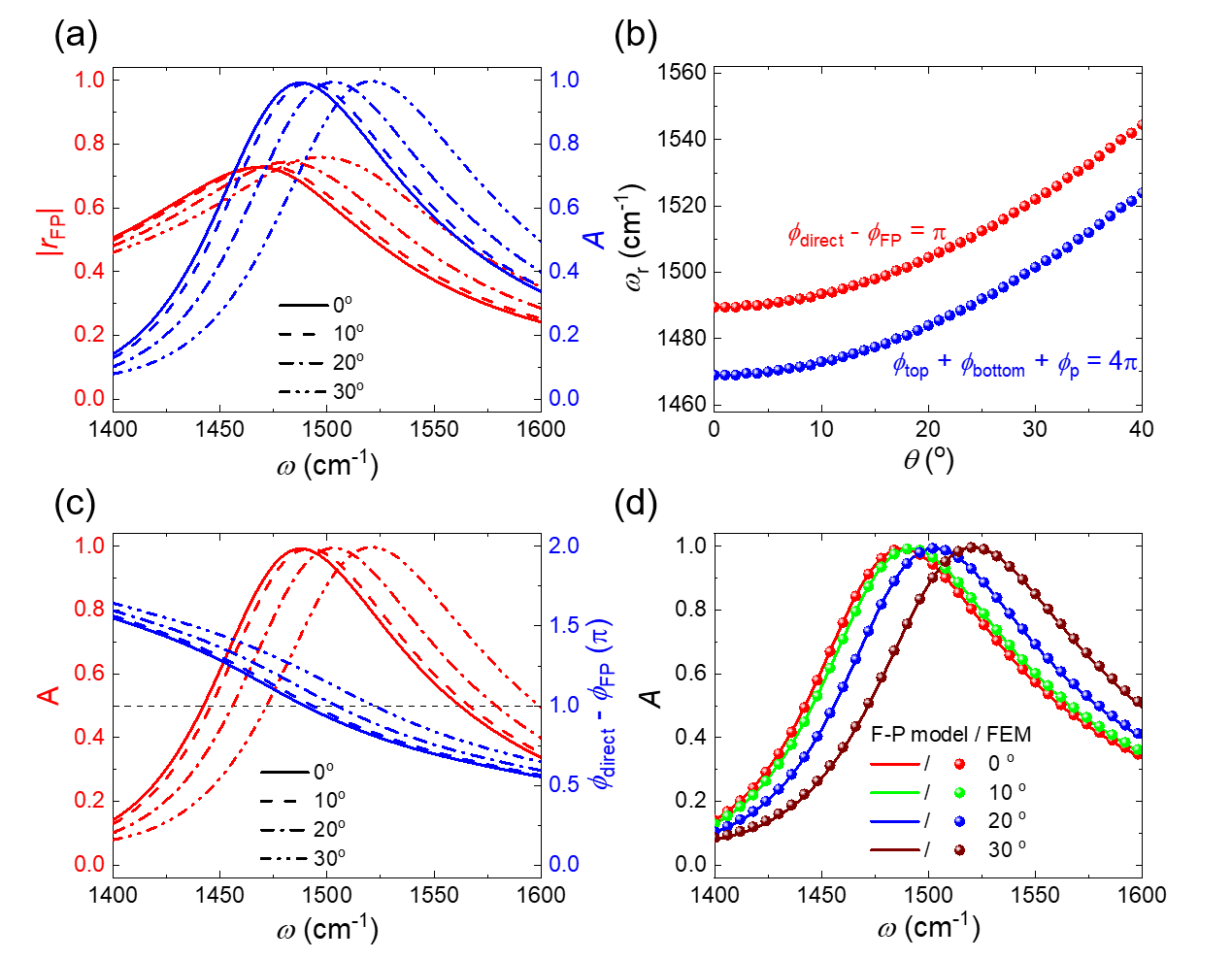}
\centering
\caption*{\textbf{Fig. S10. Analysis of incident angle dependence using the semi-analytical model.} (a) The amplitude of $\rFP$ (left axis) and total absorption derived from F-P model (right axis) for different incident angles. (b) The resonance frequencies $\omr$ and $\omFP$ as a function of incident angles. (c) The F-P model total absorption (left axis) and phase difference (right axis) for different Fermi levels of graphene. (d) The total absorption from FEM-based full wave simulation and F-P model.
\vspace{-2mm}}
\end{figure*}

We note that the maximum absorption phase condition, $\pd - \pFP = \pi$, is worked when the amplitude of resonant mode is comparable with non-resonant direct reflection at the resonance frequency. The strength of F-P resonance is inversely proportional to the Fermi level due to increased free carrier absorption in the graphene (Fig. S11(a)). Therefore, this assumption cannot be satisfied at a high Fermi level, and the difference between resonance frequencies, $\omr$, derived from the phase condition and the model-based calculation becomes larger as the increase of Fermi level (Fig. S11(b)). On the other hand, the maximum absorption phase condition is worked regardless of the incident angle due to the conservation of F-P resonance strength (Fig. S11(c)).

\begin{figure*}[t]
\includegraphics[width = 1.0\linewidth]{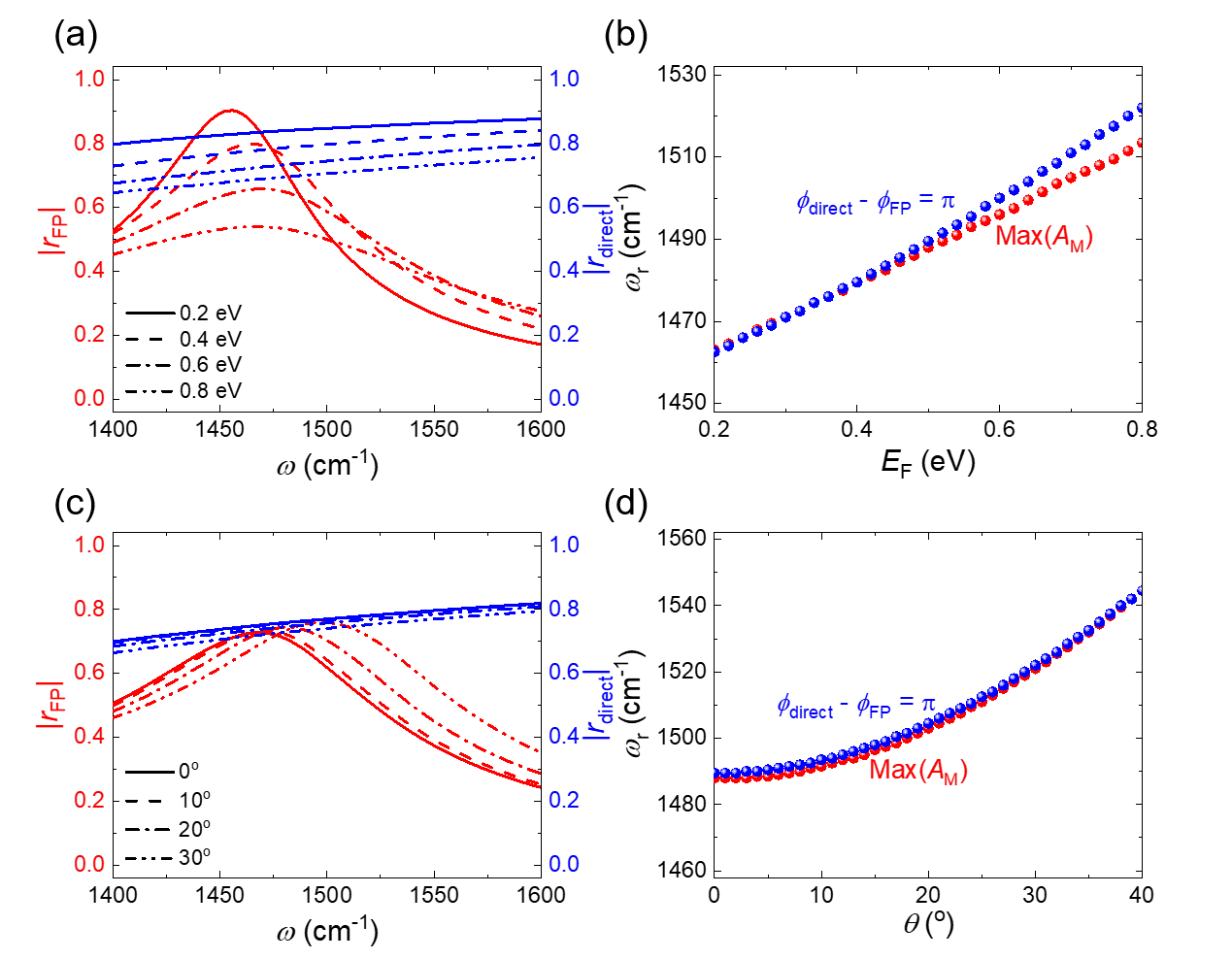}
\centering
\caption*{\textbf{Fig. S11. Range of applicability for the semi-analytical model: Fermi level and incident angle.} The amplitude of $\rFP$ and $\rd$ for different (a) Fermi levels of graphene and (c) incident angles. The resonance frequencies $\omr$ and $\omFP$ as a function of (b) Fermi levels and (d) incident angles.
\vspace{-2mm}}
\end{figure*}

Calculation of emission angle at measured frequency is required to obtain desired functionality of the proposed device. However, this process requires the calculation of angular spectra for broad frequency spectrum and various Fermi levels. Instead of finite element method(FEM)-based full-wave angular spectrum calculation, we try to obtain angular spectrum from reflection and transmission coefficients derived from the surface admittance model. The Fresnel coefficients of the graphene metasurface are described by the following equations:
\begin{equation}
    r = \frac{\tilde{{Y}}_\TU{i}\TU{cos}\theta_\TU{t} -\tilde{{Y}}_\TU{t}\TU{cos}\theta_\TU{i} - \tilde{{Y}}_\TU{s,r}\TU{cos}\theta_\TU{i}\TU{cos}\theta_\TU{t}}{\tilde{{Y}}_\TU{t}\TU{cos}\theta_\TU{i} + \tilde{{Y}}_\TU{i}\TU{cos}\theta_\TU{t} + \tilde{{Y}}_\TU{s,r}\TU{cos}\theta_\TU{i}\TU{cos}\theta_\TU{t}}
\end{equation}

\begin{equation}
    t = \frac{2\tilde{{Y}}_\TU{i}\TU{cos}\theta_\TU{i}}{\tilde{{Y}}_\TU{t}\TU{cos}\theta_\TU{i} + \tilde{{Y}}_\TU{i}\TU{cos}\theta_\TU{t} + \tilde{{Y}}_\TU{s,t}\TU{cos}\theta_\TU{i}\TU{cos}\theta_\TU{t}}
    \vspace{5mm}
\end{equation}
The calculated total absorption derived from these coefficients agrees well with the absorption calculated using FEM, as depicted in Fig. S12(a).

\begin{figure*}[t]
\includegraphics[width = 1.0\linewidth]{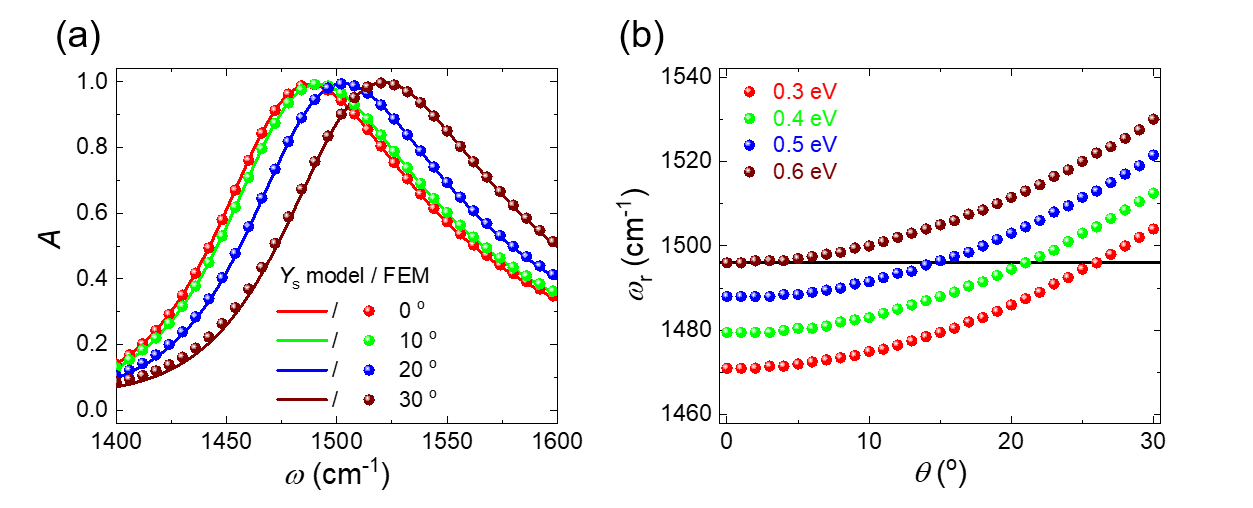}
\caption*{\textbf{Fig. S12. Validation of the surface admittance-based F-P model.} (a) The total absorption from FEM-based full wave simulation and surface admittance-based F-P model. (b) The resonance frequency $\omr$ as a function of incident angles for various Fermi levels of graphene. The black line indicates the resonance frequency for $\EF$ = \SI{0.6}{eV}. 
\vspace{-2mm}}
\end{figure*}

As a further step, we developed a graphical method that enables the rapid identification of required Fermi levels to achieve the desired emission angle at a given operating frequency. Figure S12(b) illustrates the calculated resonance frequencies $\omr$ of the device as a function of the incident angle of light. In this plot, we include a straight line with the measurement frequency as the y-intercept. The intersection point of the resonance frequency curve and the straight line indicates the emission angle at the measurement frequency. This approach allows for the rapid estimation of the emission angle, which is essential for thermal emission steerer design.

\pagebreak

\section*{Supplementary Note 4. Higher order Fabry-Perot resonance peak}

\begin{figure*}[h]
\includegraphics[width = 1.0\linewidth]{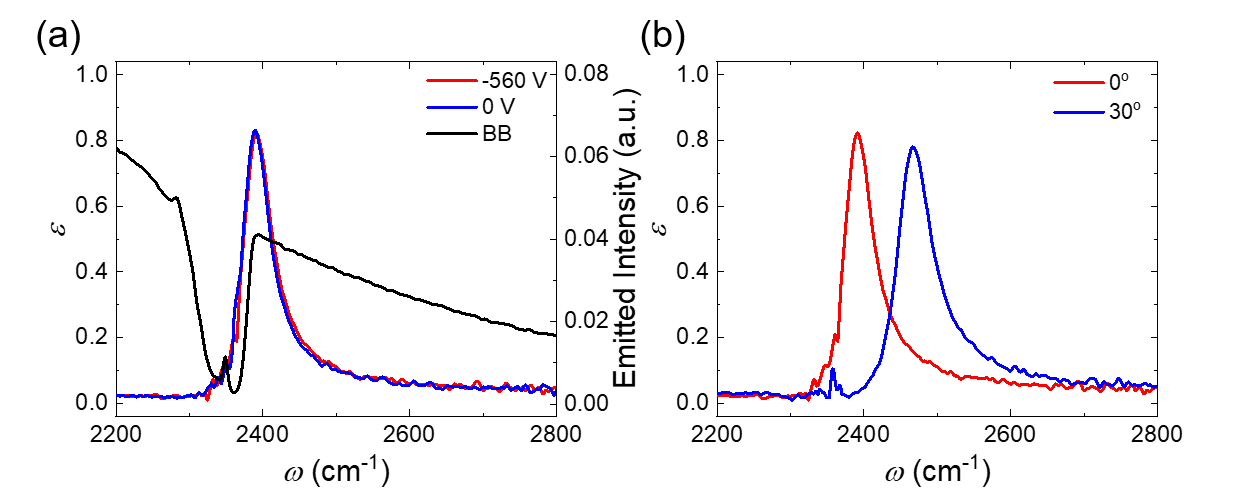}
\centering
\caption*{\textbf{Fig. S13. Emission spectra in the high frequency regime} The measured emission spectrum in high frequency regime as a function of (a) applied voltages and (b) incident angles.
\vspace{-2mm}}
\end{figure*}
\pagebreak

\section*{Supplementary Note 5. Analysis of potential factors affecting device performance}
\subsection*{A. Elemental absorption analysis}

\begin{figure*}[h]
\includegraphics[width = 1.0\linewidth]{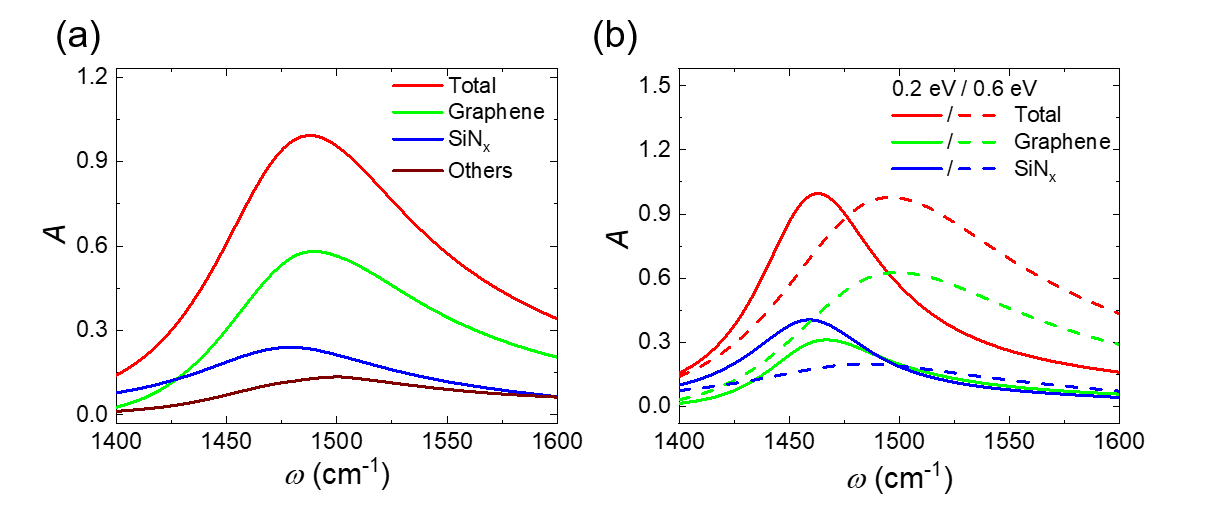}
\centering
\caption*{\textbf{Fig. S14. Elemental absorption analysis of the device.} (a) The absorbed power by different material elements with $\EF$ = \SI{0.5}{eV}. (b) The total, graphene and SiN$_{x}$ absorption for $\EF$ = \SI{0.2}{eV} and $\EF$ = \SI{0.6}{eV}.
\vspace{-2mm}}
\end{figure*}

To deeply understand the resonance behavior in the F-P resonator with a non-resonant metasurface, we performed FEM simulation to calculate elemental absorptions. The absorption of elements was calculated from $\frac{1}{P_0}\int_S \frac{1}{2}\TU{Re}(\textbf{J} \cdot \textbf{E}^*)dS$ for graphene sheet and $\frac{1}{P_0}\int_S \frac{\omega}{2}\TU{Im}(\epsilon_c)|\textbf{E}|^2dS$ for finite thickness films where $P_0$, $\textbf{J}$, $\textbf{E}$, and $\epsilon_c$ are incident wave power, current density, electric field, and complex permittivity of materials. Figure S14(a) shows the elemental absorption of Graphene, SiN, and other components. It is important to emphasize that the dominant absorption occurs in the graphene sheet and SiN$_\TU{x}$ membrane. Consequently, the total absorption peak is formed by the summation of these two absorption components. As a result, the resonance peak of the total absorption exhibits a broader frequency range and a larger shift than a single F-P resonance peak, as illustrated in Fig. S14(b).

\pagebreak

\subsection*{B. Geometry and material parameters}
We analyze the impact of material and geometry on the device caused by imperfect fabrication. Figure S15(a) shows the total absorption of the device for various geometric parameters. The deviation in structural parameters was determined by considering the fabrication tolerance specific to each fabrication process. Among the geometrical parameters, the gap width, slit width, and HfO$_{2}$ thickness exhibit noticeable resonance peak frequency shifts and broadening. This is because electromagnetic energy density at the surface of graphene is influenced by two geometric factors: exposed graphene area (gap width/slit width) and metal-graphene distance (HfO$_{2}$ thickness). Even a \SI{2}{nm} thickness variation in HfO$_{2}$ thickness considerably alters the modulation performance of resonance frequency. The broadening of the resonance peak is directly proportional to the Fermi level due to enhanced free carrier absorption. In contrast, variation in the thickness of other elements (Al$_{2}$O$_{3}$, SiO$_\TU{x}$, and Slit) have negligible effects on the optical properties of the device. 

To investigate the impact of deviations in material optical properties, we performed calculations of the total absorption for different carrier mobilities (graphene) and permittivities (Al${2}$O${3}$, SiO$_\TU{x}$, and HfO${2}$). Figure S16(a) shows that at moderate carrier mobilities ($> \SI{300} \muU$), the resonance peak of the total absorption exhibited tolerance to deviations. However, excessively low carrier mobilities broaden the resonance peak and decrease the resonance frequency shift for Fermi level modulation. Considering the potential damage induced by fabrication processes such as dielectric deposition and e-beam exposure, the lower modulation performance observed in the fabricated devices of this project could be attributed to this effect. On the other hand, for calculating the total absorption spectra of the device for different frequency-dependent permittivity $K\epsilon_\textup{r}$, where $K$ is the scaling factor, it was observed that there is a small resonance frequency shift with a slight change in broadness for high permittivity deviations. Therefore, in the proposed scheme, the deviation in material properties of dielectric layers has minimal effect on the calculation of the total absorption.

\begin{figure*}[p]
\includegraphics[width = 1.0\linewidth]{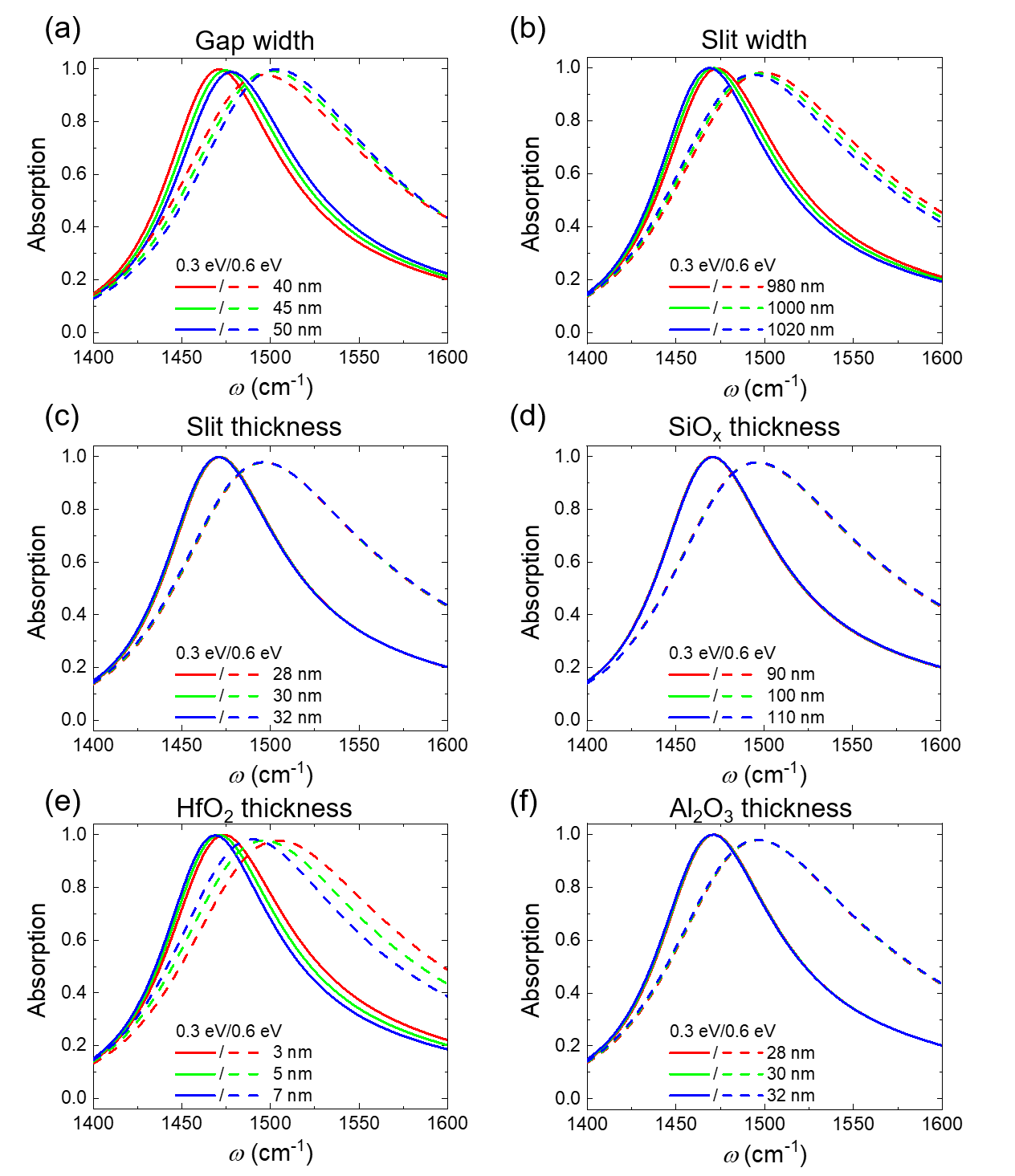}
\centering
\caption*{\textbf{Fig. S15. Deviation analysis of structural parameters in the device.} The total absorption for different geometric parameters (a) gap width, (b) slit width, (c) slit thickness, (d) SiO$_\TU{x}$ thickness, (e) HfO$_{2}$ thickness, and (f) Al$_{2}$O$_{3}$ thickness with $\EF$ = \SI{0.3}{eV} and \SI{0.6}{eV}, respectively.
\vspace{-2mm}}
\end{figure*}

\begin{figure*}[p]
\includegraphics[width = 1.0\linewidth]{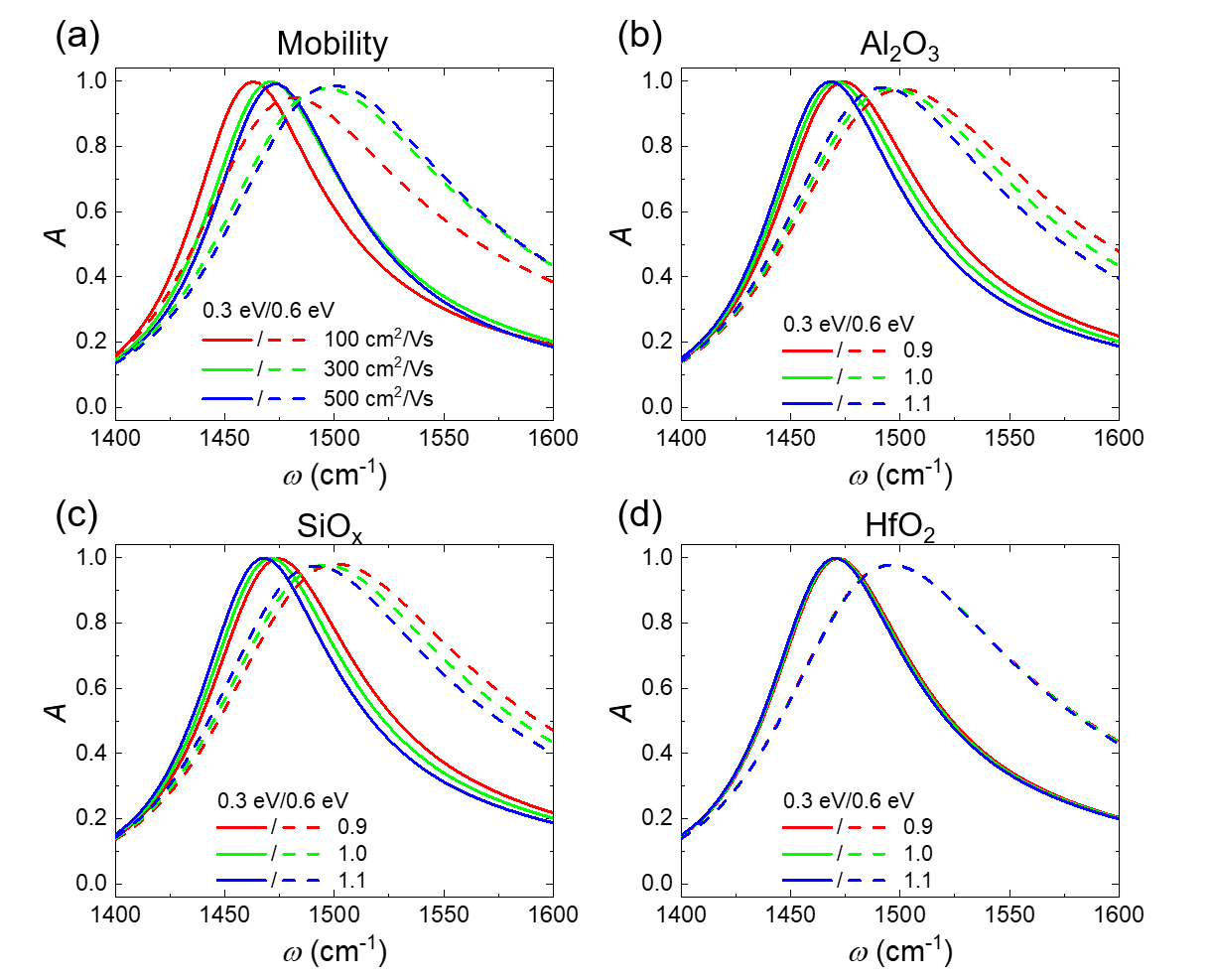}
\centering
\caption*{\textbf{Fig. S16. Deviation analysis of optical properties of materials in the device.} The total absorption for different material parameters (a) carrier mobility of graphene, the real part of permittivity of (b) Al$_{2}$O$_{3}$, (c) SiO$_\TU{x}$, and (d) HfO$_{2}$ with $\EF$ = \SI{0.3}{eV} and \SI{0.6}{eV} for different scaling factor $K$, respectively.
\vspace{-2mm}}
\end{figure*}

\pagebreak

\subsection*{C. Nonclassical effects in metal-graphene interaction}

\begin{figure*}[h]
\includegraphics[width = 1.0\linewidth]{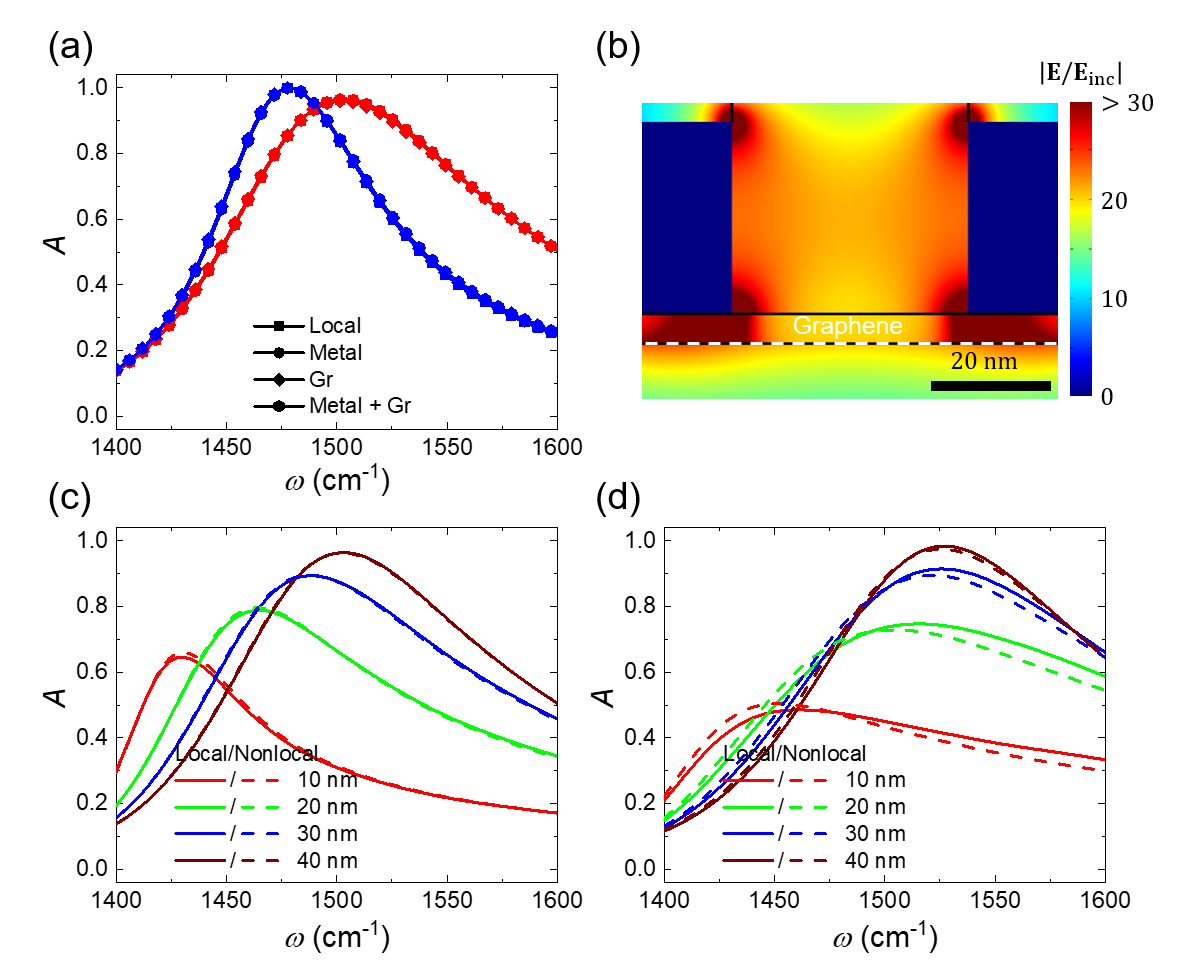}
\centering
\caption*{\textbf{Fig. S17. Analysis of nonlocal effect in the device.} (a) The total absorption for different models with $\EF$ = \SI{-0.38}{eV} (blue) and \SI{-0.68}{eV} (red). (b) The normalized magnitude of electric field distribution at resonance frequency with $\EF$ = \SI{-0.68}{eV}. The white dot line is zero thickness graphene layer. The total absorption of local and nonlocal models for different gap sizes with (c) \SI{5}{nm} and (d) \SI{1}{nm} of HfO$_{2}$ thickness.
\vspace{-2mm}}
\end{figure*}

For the proposed structure, the length scale of metallic structures (graphene and slit array) is close to the electron wavelength in materials. Therefore, nonclassical effects that are not considered in classical electromagnetic simulations, such as the nonlocal effect, quantum tunneling effect, and quantum confinement effect, need to be taken into account. Additionally, the \SI{30}{nm} thickness of metal slit is thick enough to avoid optical property change by the quantum confinement effect\cite{Quantum_size_effect}. In the case of the quantum tunneling effect between graphene-metal structures, it becomes observable for gap distance less than \SI{1}{nm} \cite{Quantum_tunneling_effect}. Thus, \SI{5}{nm} passivation HfO$_{2}$ layer between the metal slit array and graphene does not enter the quantum tunneling regime. Therefore, the nonclassical effect of concern is the nonloal effect, which has been observed in metallic structures on the scale of a few tens of nanometers. This effect arises from the nonideal spatial concentration of electrons due to quantum pressure in electron wave functions.

To investigate the nonlocal effect, we employ a hydrodynamic model for Au slit array and graphene sheet. The currents density $\textbf{J}$ inside metal and graphene sheet induced by electric field $\textbf{E}$ with frequency $\omega$ can be described by the following equation in the nonlocal frame\cite{Smith_ultimate_limits, Boardman_surface_mode}:
\vspace{5mm}
\begin{equation}
    \beta^2\nabla(\nabla\cdot \textbf{J}) + (\omega^2+i\gamma\omega) \textbf{J}) = i\omega\omega^2_p\epsilon_0\textbf{E}
    \vspace{5mm}
\end{equation}
where $\epsilon_0$, $\gamma$ and $\omega_{p}$ are the vacuum permittivity, damping coefficient and plasma frequency, respectively. The nonlocal parameter, $\beta$, depends on the Fermi level and dimensionality. We solve the equation using PDE and wave optics modules in COMSOL commericial FEM software. Figure S17(a) compares the total absorption of the device for different simulation configurations: full local, graphene nonlocal, metal nonlocal and full nonlocal. The results indicate that the fabricated device is rarely affected by the nonlocal effect. This can be attributed to two factors: (1) non-resonant scattering of the metal slit array (2) the increased effective gap width due to electric field spreading. However, when the thickness of HfO$_{2}$ is decreased to \SI{1}{nm}, the nonlocal effect starts to affect the resonance peak due to increased electric field confinement in the gap region.
\pagebreak

\subsection*{D. Leakage current in dielectric spacer}
In the classical capacitor model, the assumption in calculating induced charge density is that the applied voltage is the same value as the voltage drop along the dielectric films sandwiched by the two electrode pairs. If the voltage drops across other parts of the device are significant, the estimated induced charge density by the capacitor model may be higher than the actual charge density during the operation of the device. Thus, the inefficient gating of graphene by this factor diminishes the angle steering range and the associated spectral peak shift. To investigate the degradation of the device operation caused by inefficient gating, we measured the leakage current in the SiN$_\TU{x}$/Al$_2$O$_3$ layer at \SI{250}{\degreeCelsius}, shown in Fig. S18. The measured resistance across the dielectric spacer ($> 10^2$ M\unit{\ohm}) is many orders of magnitude higher than any electrical contact in our circuit and the resistance of the graphene sheet ($100 \sim 300$ \unit{\ohm}). Consequently, the gate voltage drop occurs almost entirely across the SiN$_\TU{x}$/Al$_2$O$_3$, despite the small leakage current.

\begin{figure*}[h]
\includegraphics[width = 0.7\linewidth]{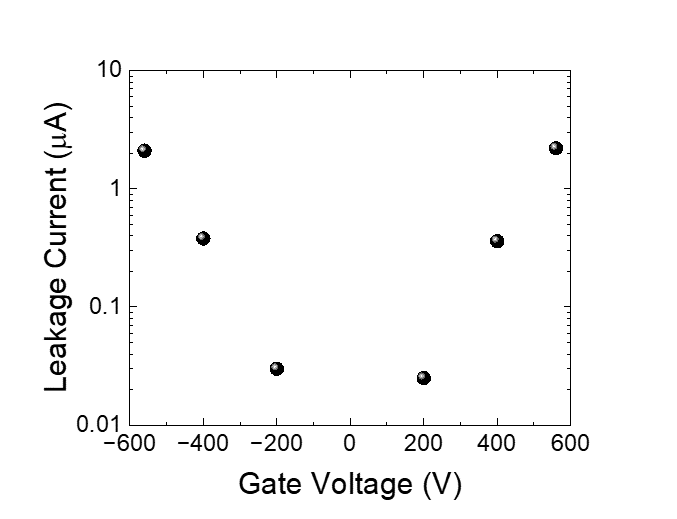}
\centering
\caption*{\textbf{Fig. S18.} Measurement of leakage current across the SiN$_\TU{x}$/Al$_2$O$_3$ layers between the graphene sheet and the bottom Au backgate. The device temperature is maintained at \SI{250}{\degreeCelsius} during measurement.
\vspace{-2mm}}
\end{figure*}

\pagebreak

\subsection*{E. Dependence of gate capacitance on temperature and gate voltage}
The capacitance between the graphene and the backgate is determined by the static dielectric constants of the dielectric materials. The electrostatic dielectric constant of the membrane layer could vary with temperature and gating voltage. Therefore, it is important to investigate the effects of the aforementioned factors to understand the discrepancy of emissivity spectra in measurement and calculation. For the SiN$_\TU{x}$ substrates used in this work, measurements in our previous study have shown that the dielectric constant increases only slightly with temperature up to \SI{250}{\degreeCelsius}, and remains unchanged by the applied gate voltage\cite{Graphene_thermal_modulation_Brar}. These measurements justify the basic capacitance model used in estimating how carrier density changes with applied gate voltage, which assumes the dielectric constants of the SiN$_\TU{x}$/Al$_2$O$_3$ to be independent of temperature and gate voltage. 

\subsection*{F. Uncertainty of dielectric constant}
Commercially available products have a variance of dielectric constant owing to the fluctuation of growth conditions. In calculating induced carrier density at the surface of graphene by gating, the dielectric constant of the SiNx memebrane layer was assumed to be 7.5, as derived in our previous work\cite{Salisbury_Jang}. However, the dielectric constant of the purchased product has a variance of 1, corresponding to $\sim$$15\%$ of dielectric constant. Considering this variance, the magnitude of the induced carrier density could be reduced to $20\%$ of our expectation.

\subsection*{G. Impurity and charge trap effect}
One of the important issues in the graphene-based active metasurface is charge traps and atmospheric impurities on or in the SiN$_\TU{x}$/Al$_2$O$_3$ which are known to change their charge state depending on the applied gate voltage. The effects of such impurities lead to deviations from the simple capacitor model. Our evidence for such impurity states is the hysteresis observed in resistance vs. gate voltage measurements, consistent with previous studies that systematically investigated such charge traps and impurities.

\pagebreak

\subsection*{F. Effect of initial doping of angle steering}

\begin{figure*}[h]
\includegraphics[width = 0.7\linewidth]{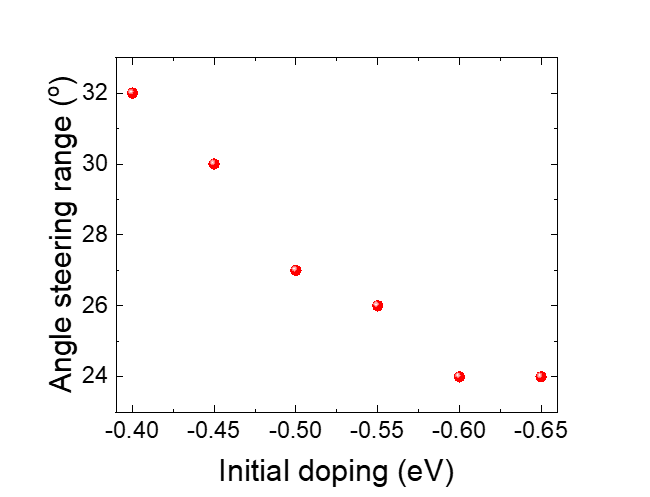}
\centering
\caption*{\textbf{Fig. S19.} Calculated angle steering range as a function of the initial doping level of graphene
\vspace{-2mm}}
\end{figure*}

The Fermi level-angular (Fig. 2(d)) and frequency-angular (Fig. S21) emissivity spectra indicate variations in both the steering angle range and modulation depth depending on the initial doping level of graphene. The correlation between the Fermi level and carrier density $n_\TU{carrier}$ follows the relationship $E_F \propto \sqrt{n_\TU{carrier}}$, suggesting that higher Fermi levels undergo less change with the same increase in carrier density (proportional to $V_G$). Consequently, a higher initial doping level results in less alteration of the Fermi level for a given applied gate voltage. Figure S19 shows that the steering angle range for $E_F = \SI{-0.4}{eV}$ at $V_G = 0$ is 6$^\circ$ higher than in the case of an initial doping of \SI{-0.55}{eV}.

\pagebreak

\section*{Supplementary Note 6. Potential angle steering capability}

\begin{figure*}[h]
\includegraphics[width = 1.0\linewidth]{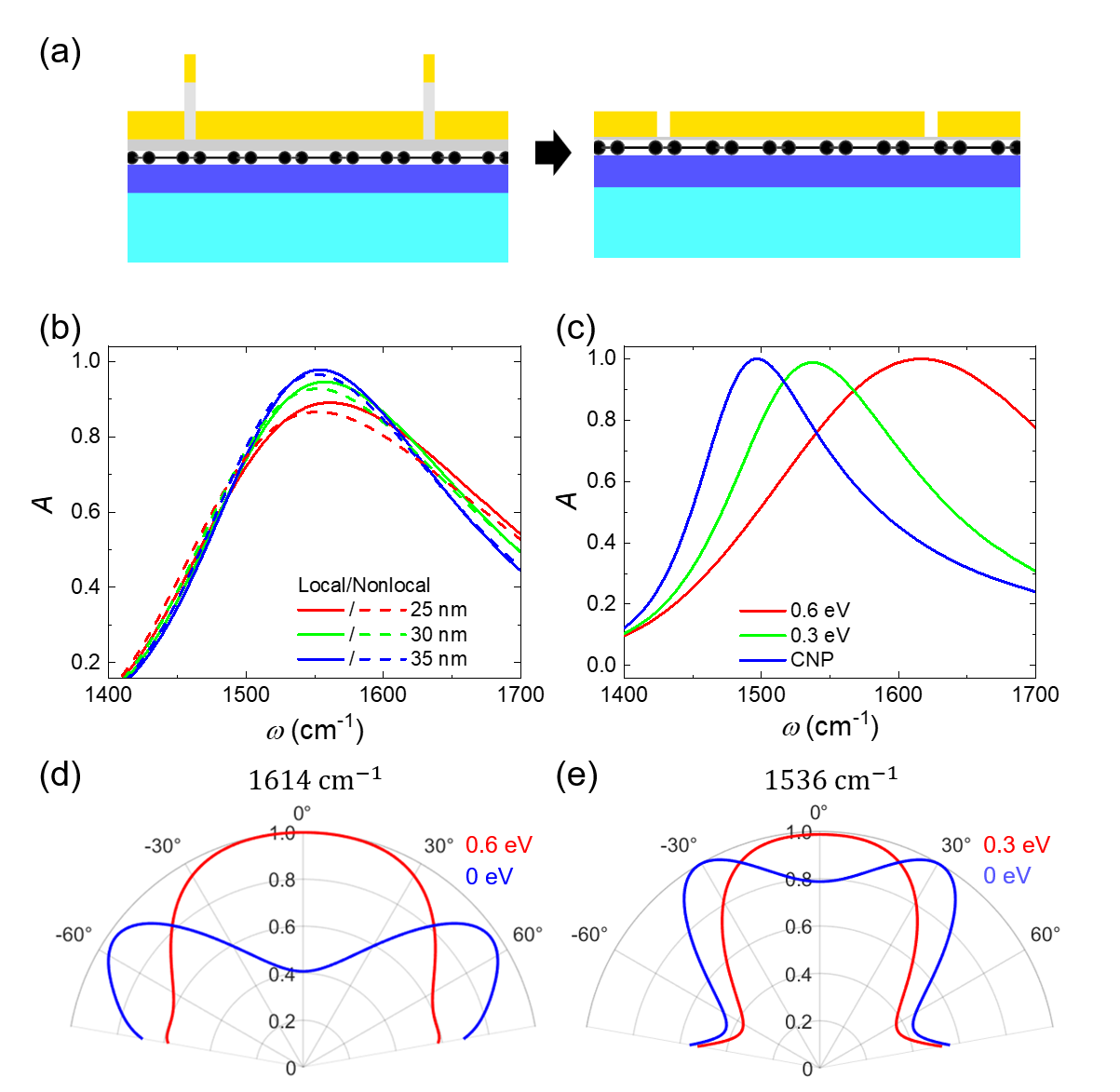}
\caption*{\textbf{Fig. S20. Theoretical performance limit of the proposed configuration.} (a) The schematic of geometry modification for optimization. (b) The total absorption of optimized structure ideal structure for different gap sizes and models with $\EF$ = \SI{0.6}{eV}. (c) The total absorption of optimized for various Fermi levels of graphene. The angular absorption spectrum for (d) \SI{0}{eV}/\SI{0.6}{eV} at \SI{1614}{cm^{-1}} and (e) \SI{0}{eV}/\SI{0.3}{eV} at \SI{1536}{cm^{-1}}.
\vspace{-2mm}}
\end{figure*}

In this study, We primarily focus on demonstrating dynamic control of directional emission angles, and the change of emission angles of the fabricated device was limited to \ang{16}. However, the proposed design scheme has the potential for a much larger emission angle change through the optimization of structural parameters and slight modifications to the configuration, as shown in Figure S20(a). The results of the structural parameter effect analysis indicate that the strength of interaction between graphene and the metal slit array strongly influences the emission angle $\theta$. To maximize electromagnetic field intensity at the surface of graphene, we propose ideal structure where HfO$_2$ layer, located on top of graphene, is reduced to \SI{1}{nm}. This configuration ensures the smallest distance between graphene and the metal slit array without the quantum tunneling effect. The gap and slit widths are optimized for the largest emission angle change. 

Considering the significant degradation of performance due to the nonlocal effect for gaps less than \SI{30}{nm}, the minimum gap width is limited to \SI{30}{nm}. For Fermi level modulation at the \SI{0}{eV} and \SI{0.6}{eV}, we obtain an emission angle change of approximately \ang{60} for gap and slit widths of \SI{30}{nm} and \SI{740}{nm}, respectively. Here, we focus on maximizing angle change. However, if the goal is to achieve emission steering with narrow beam, we can obtain a narrower beam by setting a high Fermi level at \SI{0.3}{eV}, as shown in Fig. S20(e). In this case, the maximum emission angle change is \ang{40} due to reduced Fermi level modulation. Additionally, by employing other materials with smaller material loss than graphene, we anticipate the possibility of even narrower beam steering.
\pagebreak

\section*{Supplementary Note 7. Modulation of angle-frequency spectra}

\begin{figure*}[h]
\includegraphics[width = 1.0\linewidth]{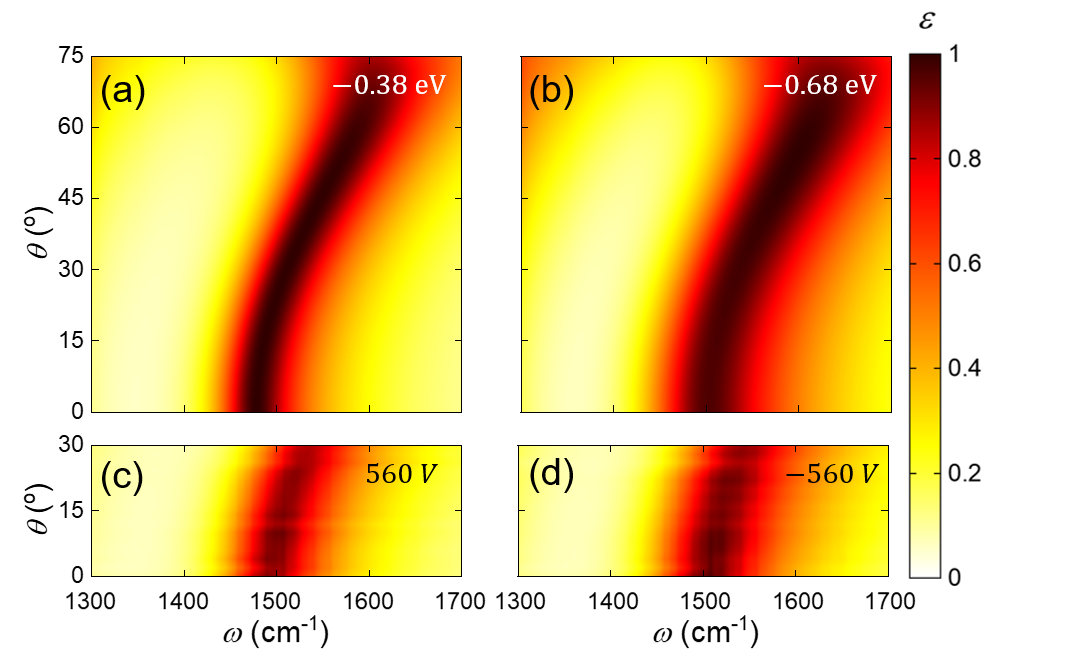}
\centering
\caption*{\textbf{Fig. S21. Comparing calculated and measured emissivity map of the device.} The calculated angle-frequency total emissivity spectra of the fabricated device for (a) $\EF$ = \SI{-0.38}{eV} and (b) \SI{-0.68}{eV}, respectively. Panels (c) and (d) show the measured angle-frequency emissivity spectra of the device for $V_G = 560$ \unit{V} and $-560$ \unit{V}, corresponding to each Fermi level.
\vspace{-2mm}}
\end{figure*}

To gain a deeper understanding of the device's operation, it is essential to analyze the emissivity behavior for the angle-frequency spectrum with fixed Fermi levels. Figures 21(a) and 21(b) depict the angle- and frequency-dependent emissivity spectra for $\EF$ = \SI{-0.38}{eV} and \SI{-0.68}{eV} corresponding to $V_G = 560$ \unit{V} and $-560$ \unit{V}, respectively. The resonance frequency gradually shifts to a higher frequency with an increase in the incident angle, consistent with the previous result. The incident angle-dependent propagation phase accumulation is compensated by frequency-dependent phase change. It is noteworthy that the change in the Fermi level of graphene induces a simple translation of the resonance frequency curve. This constant frequency translation of the curve is inferred through nonresonant phase modulation of the graphene metasurface. The width of the resonance peak is broader for a higher Fermi level due to increased optical loss (see Supplementary Note 5). The experimentally measured angle- and frequency-dependent emissivity is well matched with the calculation results, as shown in Fig. S21(c) and (d).

\pagebreak